\documentclass[12pt,preprint]{aastex}
\usepackage{emulateapj5}
\usepackage{apjfonts}
\usepackage{epsfig}
\usepackage{natbib}
\newcommand{\cat}{CaT}
\newcommand{\chk}{Ca~H+K}
\newcommand{\chisq}{\ensuremath{\chi^2}}

\newcommand{\delm}{\ensuremath{\Delta M_\mathrm{BH}}}

\def\gtrsim{\mathrel{\hbox{\rlap{\hbox{\lower4pt\hbox{$\sim$}}}\hbox{\raise2pt\hbox{$>$}}}}}

\newcommand{\galfit}{\ensuremath{\tt GALFIT}}

\newcommand{\halpha}{H\ensuremath{\alpha}}

\newcommand{\hbeta}{H\ensuremath{\beta}}

\newcommand{\hst}{\emph{HST}}

\newcommand{\kms}{km~s\ensuremath{^{-1}}}

\newcommand{\lledd}{\ensuremath{L_{\mathrm{bol}}/L{\mathrm{_{Edd}}}}}

\newcommand{\loiii}{\ensuremath{L_{\mathrm{[O {\tiny III}]}}}}

\newcommand{\mbulge}{\ensuremath{M_\mathrm{BH}-M_{\mathrm{bulge}}}}

\newcommand{\ml}{\ensuremath{\Upsilon}}

\newcommand{\mlr}{\ensuremath{\Upsilon_r}}

\newcommand{\mbh}{\ensuremath{M_\mathrm{BH}}}

\newcommand{\mgb}{\ion{Mg}{1}$b$}
\newcommand{\mlb}{\ensuremath{M_{\mathrm{BH}}-L_{\mathrm{bulge}}}}

\newcommand{\msigma}{\ensuremath{M_{\mathrm{BH}}-\sigmastar}}
\newcommand{\msun}{\ensuremath{M_{\odot}}}

\newcommand{\nii}{[\ion{N}{2}]}

\newcommand{\oiii}{[\ion{O}{3}]}

\newcommand{\re}{\ensuremath{r_{\mathrm e}}}

\newcommand{\sers}{S{\'e}rsic}

\newcommand{\sii}{[\ion{S}{2}]}

\newcommand{\sigmastar}{\ensuremath{\sigma_{\ast}}}

\newcommand{\xmm}{{\it XMM-Newton}}

\def\lax{{$\mathrel{\hbox{\rlap{\hbox{\lower4pt\hbox{$\sim$}}}\hbox{$<$}}}$}}
\def\gax{{$\mathrel{\hbox{\rlap{\hbox{\lower4pt\hbox{$\sim$}}}\hbox{$>$}}}$}}

\slugcomment{Draft version 6; June 3, 2010;
to be submitted to {\it The Astrophysical Journal}.}
\shorttitle{{\it Megamaser Hosts}}
\shortauthors{GREENE ET AL.}

\begin{document}

\title{Precise Black Hole Masses From Megamaser Disks: 
Black Hole-Bulge Relations at Low Mass}

\author{Jenny E. Greene}
\affil{Department of Astrophysical Sciences, Princeton University, 
Princeton, NJ 08544; Princeton-Carnegie Fellow}

\author{Chien Y. Peng}
\affil{NRC Herzberg Institute of Astrophysics, 5071 West Saanich Road, 
Victoria, BC V9E2E7, Canada}

\author{Minjin Kim, Cheng-Yu Kuo, James A. Braatz, C.~M.~Violette Impellizzeri, 
James J. Condon, K. Y. Lo}

\affil{National Radio Astronomy Observatory, 520 Edgemont Road, Charlottesville, VA, USA}

\author{Christian Henkel}

\affil{Max Planck Institute for Radio Astronomy, Auf dem Hugel, Bonn, Germany}

\author{Mark J. Reid}

\affil{Harvard-Smithsonian Center for Astrophysics, 60 Garden Street, 
Cambridge, MA 02138, USA}

\begin{abstract}

The black hole (BH)-bulge correlations have greatly influenced the
last decade of effort to understand galaxy evolution. Current
knowledge of these correlations is limited predominantly to high BH
masses (\mbh$\gtrsim 10^8$~\msun) that can be measured using direct
stellar, gas, and maser kinematics. These objects, however, do not
represent the demographics of more typical $L< L^*$ galaxies.  This
study transcends prior limitations to probe BHs that are an order of
magnitude lower in mass, using BH mass measurements derived from the
dynamics of H$_2$O megamasers in circumnuclear disks. The masers
trace the Keplerian rotation of circumnuclear molecular disks
starting at radii of a few tenths of a pc from the central BH.
Modeling of the rotation curves, presented by Kuo et al. (2010),
yields BH masses with exquisite precision.  We present stellar
velocity dispersion measurements for a sample of nine megamaser disk 
galaxies based
on long-slit observations using the B\&C spectrograph on the Dupont
telescope and the DIS spectrograph on the 3.5~m telescope at Apache
Point.  We also perform bulge-to-disk decomposition of a subset of
five of these galaxies with SDSS imaging.  The maser galaxies as a
group fall below the \msigma\ relation defined by elliptical
galaxies.  We show, now with very precise BH mass measurements, that
the low-scatter power-law relation between \mbh\ and \sigmastar\
seen in elliptical galaxies is not universal.  The elliptical galaxy
\msigma\ relation cannot be used to derive the BH mass function at
low mass or the zeropoint for active BH masses.  The processes
(perhaps BH self-regulation or minor merging) that operate at higher
mass have not effectively established an \msigma\ relation in this
low-mass regime.

\end{abstract}


\section{Dynamical Black Hole Masses}

The past two decades have seen substantial improvements in our
understanding of the demographics of nuclear supermassive black holes
(BHs), predominantly driven by the availability of BH mass
measurements for nearby galaxies from dynamical techniques.  The
advent of the \emph{Hubble Space Telescope} has enabled a few dozen
dynamical BH mass measurements, which reveal that supermassive BHs are
ubiquitous components of bulge-dominated galaxy centers
\citep[e.g.,][]{kormendy2004}. Furthermore, there appears to be a
remarkably tight correlation between BH mass and bulge properties,
including velocity dispersion \citep[the \msigma\ relation;
][]{gebhardtetal2000a,ferraresemerritt2000,
  tremaineetal2002,gultekinetal2009}, luminosity \citep[the \mlb\
relation, e.g.,][]{marconihunt2003}, and mass
\citep[e.g.,][]{haeringrix2004}.  The apparently tiny scatter in the
\msigma\ relation has led to a common view that the growth of BHs is
intimately connected to the growth of the surrounding galaxy
\citep[see discussion in][]{ho2004}.  In this paper, we explore the
demographics of a new sample of dynamical BH masses measured using
observations of circumnuclear masers (Kuo, C.~Y. et al. in prep.).

\begin{figure*}
\vbox{ 
\hskip 0.7in
\psfig{file=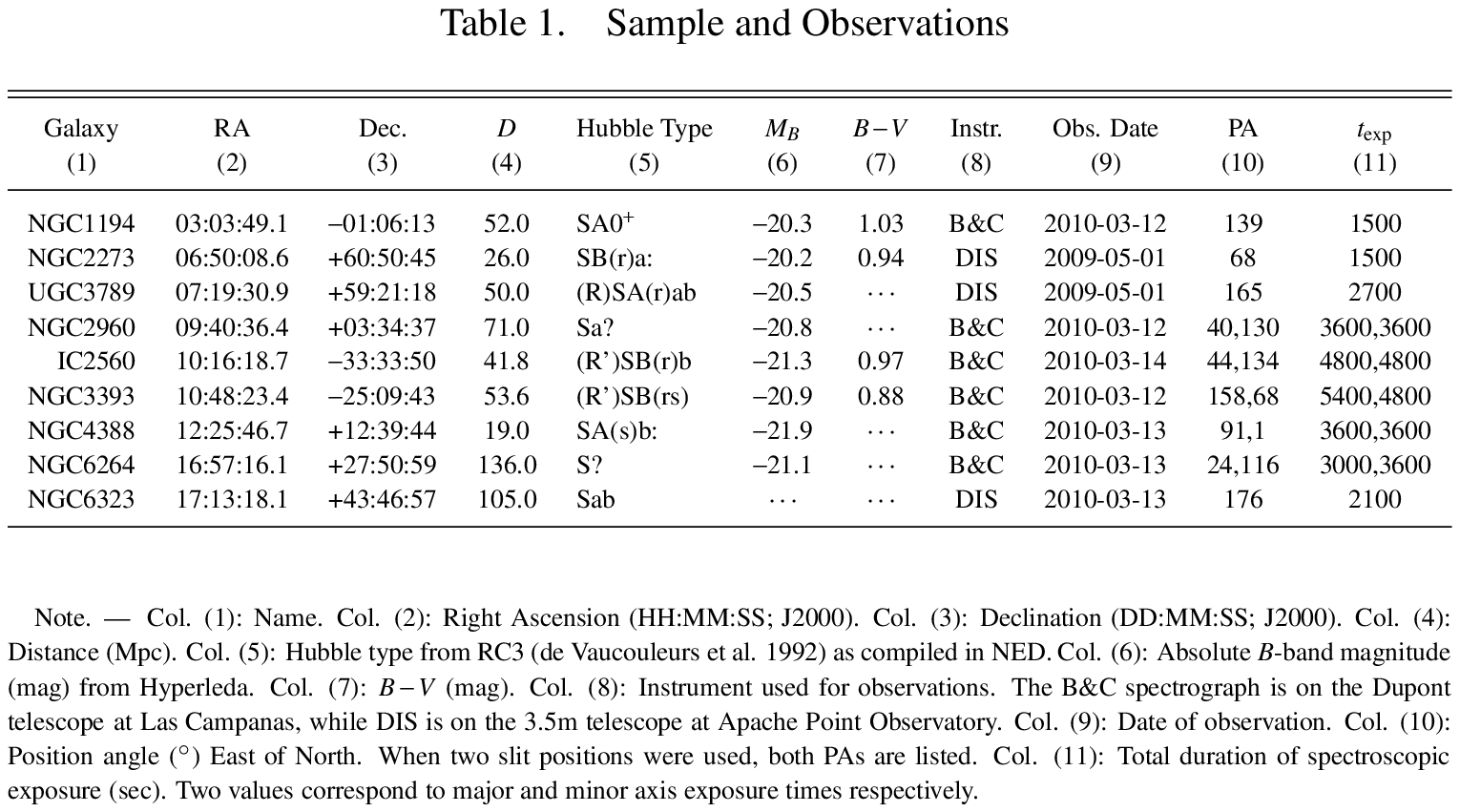,width=0.8\textwidth,keepaspectratio=true,angle=0}
}
\end{figure*}

Apart from our own Galactic Center, where individual stars are used as
test particles to weigh the BH
\citep[e.g.,][]{ghezetal2008,gillessenetal2009}, the most precise BH
masses are derived using water megamaser spots in a Keplerian
circumnuclear disk \citep[e.g.,][]{miyoshietal1995}.  Such megamasers
are very close to an ideal dynamical tracer of the central mass at the
galaxy center, as they delineate a disk with a well-measured Keplerian
rotation curve within fractions of a pc from the center.  They are
detectable when the disk is aligned within a few degrees of the line
of sight.  Masers detected in edge-on nuclear disks have
characteristic triple-peaked spectral profiles, making the disk
inclination angle well-known in these cases \citep{lo2005}.  BH mass
measurements based on the Keplerian rotation curves of circumnuclear
megamaser disks are not prone to the same systematic uncertainties as
the dynamical methods based on optical observations.  In the case of
gas dynamical methods, it can be very difficult to determine the level
of non-virial gas support (e.g.,~due to turbulence) but neglecting
these extra sources of support can lead to significant bias in the
final result \citep[e.g.,][]{barthetal2001}.  Stellar dynamical
methods are sensitive to assumptions about galaxy anisotropy and
radial variation in dark matter fraction that are not well constrained
observationally in general
\citep[e.g.,][]{gebhardtthomas2009,vandenboschdezeeuw2010}.

The cleanest and best-studied megamaser galaxy is the nearby Seyfert
galaxy NGC 4258 \citep[e.g.,][and for a complete literature review see
Lo 2005]
{nakaietal1993,greenhilletal1995,miyoshietal1995,herrnsteinetal1996}.
Observations of centripetal acceleration from the frequency change of
the maser components at the systemic velocity of the system, combined
with VLBA images of the megamaser disk, allow for the most precise
extragalactic angular diameter distance measurement
\citep{herrnsteinetal1999}.  Other observations provide detailed
constraints on the thickness, velocity structure, and toroidal
magnetic field in an accretion disk
\citep[e.g.,][]{modjazetal2005,herrnsteinetal2005,
  argonetal2007,humphreysetal2008}.  For a long time it proved
difficult to detect additional well-defined Keplerian megamaser disks,
\citep[e.g.,][]{greenhilletal1997,braatzetal1997,kondratkoetal2006a},
but the situation has changed dramatically with the advent of the
Green Bank Telescope (GBT) of the NRAO\footnote{The National Radio
  Astronomy Observatory is a facility of the National Science
  Foundation operated under cooperative agreement by Associated
  Universities, Inc.}.  Thanks to the increased sensitivity of the GBT
with a wide-band spectrometer, more megamaser candidates have been
discovered \citep[][]{braatzgugliucci2008,greenhilletal2009}.  Of the
$\sim 130$ extragalactic megamasers, approximately half were
discovered with the GBT and $\sim 20$ show evidence of originating in
a disk.

The primary goal of ongoing megamaser searches is to find additional
disk masers that can be used for angular diameter distance
measurements in the Megamaser Cosmology Project
\citep[MCP;][]{reidetal2009,braatzetal2010}.  With a sufficient number
of angular diameter distances to galaxies in the Hubble flow, the MCP
aims to determine a precise independent Hubble constant to a few
percent accuracy \citep[][]{reidetal2009,braatzetal2010}.  An
important outcome of the MCP is that the Keplerian megamaser disks
also provide a precise measurement of the enclosed mass within a few
tenths of a pc from the center of these spiral galaxies, dominated by
a BH at the center, a valuable resource in its own right.  The BH
masses for seven maser galaxies are presented in Kuo et al. (in
preparation).  The present paper examines the demographics of the new
megamaser galaxies, focusing on the shape and scatter of BH-bulge
relations for these galaxies.

To date, the vast majority of dynamical BH masses have been obtained
in massive elliptical and S0 galaxies (Fig. 1).  There are only five
targets with BH mass measurements $< 10^7$~\msun\ (one of those being
the Milky Way) and there are only 11 spiral galaxies with Hubble type
of Sa or later in the sample of 67 galaxies used in the most recent
calibration of the \msigma\ relation \citep[][
Fig. 1]{gultekinetal2009}.  The disproportionate representation of
massive galaxies is easily understood.  Local elliptical galaxies are
relatively dust-free, and the BHs are massive enough that we can
resolve their gravitational spheres of influence at distances of tens
of Mpc. Spiral galaxies tend to be dusty and star-forming,
complicating dynamical measurements.  Our knowledge of BH demographics
at low masses is rudimentary, such that the shape and scatter in
BH-bulge scaling relations as well as the BH occupation fraction in
spiral galaxies remain uncertain \citep[e.g.,][]{greeneho2007b}.  The
main focus of this work is to exploit the new sample of reliable BH
masses afforded by the new maser studies to explore BH demographics
for BHs with \mbh$\sim 10^7$~\msun\ in spiral galaxies.

The paper starts with a description of the sample (\S 2).  We then
describe the observations and data reduction (\S3) and the
spectroscopic (\S 4) and imaging (\S 5) analysis.  In \S 6 we present
the primary result of the paper, namely the location of the maser
galaxies in the \msigma\ plane, while in \S 7 the scalings between
\mbh\ and bulge luminosity/mass are briefly discussed.  The impact of
our results on our knowledge of BH demographics, particularly as
probed with active galaxies, is discussed in \S 8, and we summarize
and conclude in \S 9.

Throughout we assume the following cosmological parameters to calculate
distances: $H_0 = 70$~\kms~Mpc$^{-1}$, $\Omega_{\rm m} = 0.30$,
and $\Omega_{\Lambda} = 0.70$.

\section{The Sample and The Black Hole Masses}

Over the years there have been a number of searches for megamaser
galaxies \citep[e.g.,][]{braatzetal1997,kondratkoetal2006a,
  braatzgugliucci2008}.  By and large, the surveys have targeted known
obscured active galaxies, with a detection rate of $\sim 5\%$.  Once
\vbox{ 
\vskip 18mm
\psfig{file=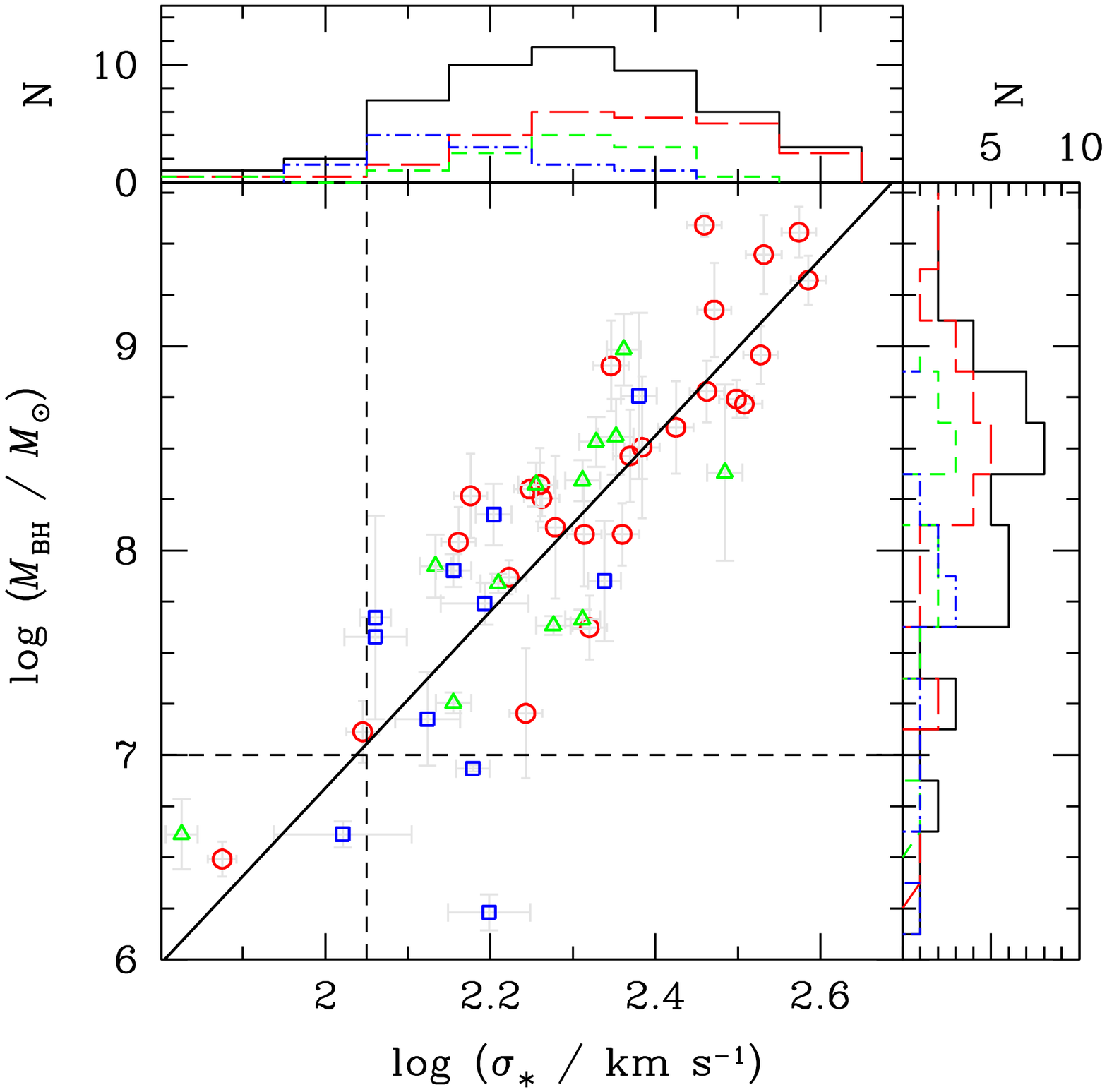,width=0.45\textwidth,keepaspectratio=true,angle=0}
}
\vskip -0mm
\figcaption[]{The \msigma\ relation as defined by nearby inactive galaxies 
from \citet{gultekinetal2009},
color-coded by morphological type of the host galaxy.  We
distinguish between elliptical ({\it red circles}), S0 ({\it green
  triangles}), and spiral ({\it blue squares}) galaxies.  The
distribution of stellar velocity dispersion is shown in the black histogram in
the top panel, also divided into the contribution of elliptical
({\it red long-dashed lines}), S0 ({\it green short-dashed lines}),
and spiral ({\it blue dot-dashed lines}) galaxies.  Similarly, the
\mbh\ distributions are shown in the right-hand panel.  The
histograms, as well as the black dashed lines, highlight 
the dearth of objects at \sigmastar$\lesssim 125$~\kms\ and 
\mbh$\lesssim 10^7$~\msun, the regime
investigated by the current work.\label{fits}}
\vskip 5mm
\noindent
detected in a single-dish survey, megamasers that show disk
characteristics, e.g., both systemic and high-velocity Doppler
components, and that are bright enough, are subsequently imaged with
Very Long Baseline Interferometry (VLBI) to map the positions and
velocities of individual maser spots.  The map determines whether a
maser is in a clean dynamical system or has more complicated
structures such as outflows \citep[e.g.,][]{kondratkoetal2005}.

In this paper we present velocity dispersions both for the Kuo et
al. sample and two other objects, NGC 3393 \citep{kondratkoetal2008}
and IC 2560 \citep{ishiharaetal2001}.  In the latter case, while it is
assumed that the systemic masers arise from the inner edge of the
masing disk, without VLBI mapping the enclosed mass is uncertain by
factors of a few.  We expect better data will yield a concrete mass.
The VLBI (and thus mass constraints) for the remaining seven objects
are presented for the first time in C-Y. Kuo et al. in preparation.
The details of the BH mass fitting are described in Kuo et al. but are
reviewed here briefly for completeness.  The BH masses are derived by
fitting a Keplerian rotation curve to the positions and velocities of
the maser spots.  The distance is fixed using an assumed value of
$H_0$ and the inclination of the maser disk is assumed to be $90
\arcdeg$.  All of the megamaser disks for which we have measured an
inclination are within $6 \arcdeg$ of being perfectly edge-on, so the
inclination contributes less than a $1\%$ error to the measurement of
the black hole mass. In NGC 4388 we could not measure a disk
inclination, but even if the disk is $20 \arcdeg$ from edge-on, the
contribution to the error in the BH mass would be only $12\%$, which
is comparable to the BH mass error caused by the distance uncertainty
($11\%$).  Ongoing monitoring of the most promising
targets will allow the measurement of accelerations in the systemic
features.  Joint modeling of \mbh\ and distance for these targets will
lead to a tighter constraint on each
\citep[e.g.,][]{herrnsteinetal1999}.

It is interesting to note that the BH masses are strongly clustered
around $\sim 10^7$~\msun.  We suspect that this is a simple
consequence of picking random galaxies from the local active-galaxy
mass function.  Without accounting for incompleteness, the observed
active mass function is strongly peaked at a mass $\sim 10^7$~\msun\
\citep[e.g.,][]{heckmanetal2004}.  At higher mass, there are very few
systems radiating at an appreciable fraction of their Eddington limit
and we are woefully incomplete at lower mass
\citep{greeneho2007b,schulzewisotzki2010}.  While we do not sample the
full range of BH masses in the local Universe with this sample, our
selection is unbiased with respect to bulge mass and velocity
dispersion and thus does not invalidate our results.

The megamaser galaxies are found predominantly in early-to-mid--type
spiral galaxies, ranging in Hubble type from S0 (e.g., NGC 1194) to Sb
(e.g., NGC 4388).  Table 1 includes the RC3 morphological types
\citep{devaucouleursetal1992} and $B$-band luminosities and colors
from Hyperleda \citep{patureletal2003}.  As noted above, the parent
sample consists predominantly of known active galaxies, ranging in
distance from 18 to 150 Mpc.  Only UGC 3789 did not have a documented
active galactic nucleus (AGN) when the circumnuclear maser was
discovered, although is it a narrow-line AGN.  With the exception of
NGC 4388, which has a weak broad \halpha\ line \citep{hfs1997broad},
the active galaxies are classified as narrow-line (obscured) AGNs from
the optical spectra.  Furthermore, those with X-ray spectra have large
column densities. Most are Compton thick \citep[$N_H >
10^{24}$~cm$^{-2}$;][]
{guainazzietal2005,madejskietal2006,zhangetal2006,
  kondratkoetal2006b,greenhilletal2008,zhangetal2010}.  The only
exceptions are NGC 4388, which has been observed to undergo large
variations in column density over time scales of hours
\citep{elvisetal2004}, and NGC 6264, NGC 6323, and UGC 3789, which
have not yet been observed in the X-rays.  The attempted \xmm\
observations of the former two were destroyed by flaring.

\section{Observations and Data Reduction}

\begin{figure*}
\vbox{ 
\vskip -3.4mm
\hskip 0.6in
\psfig{file=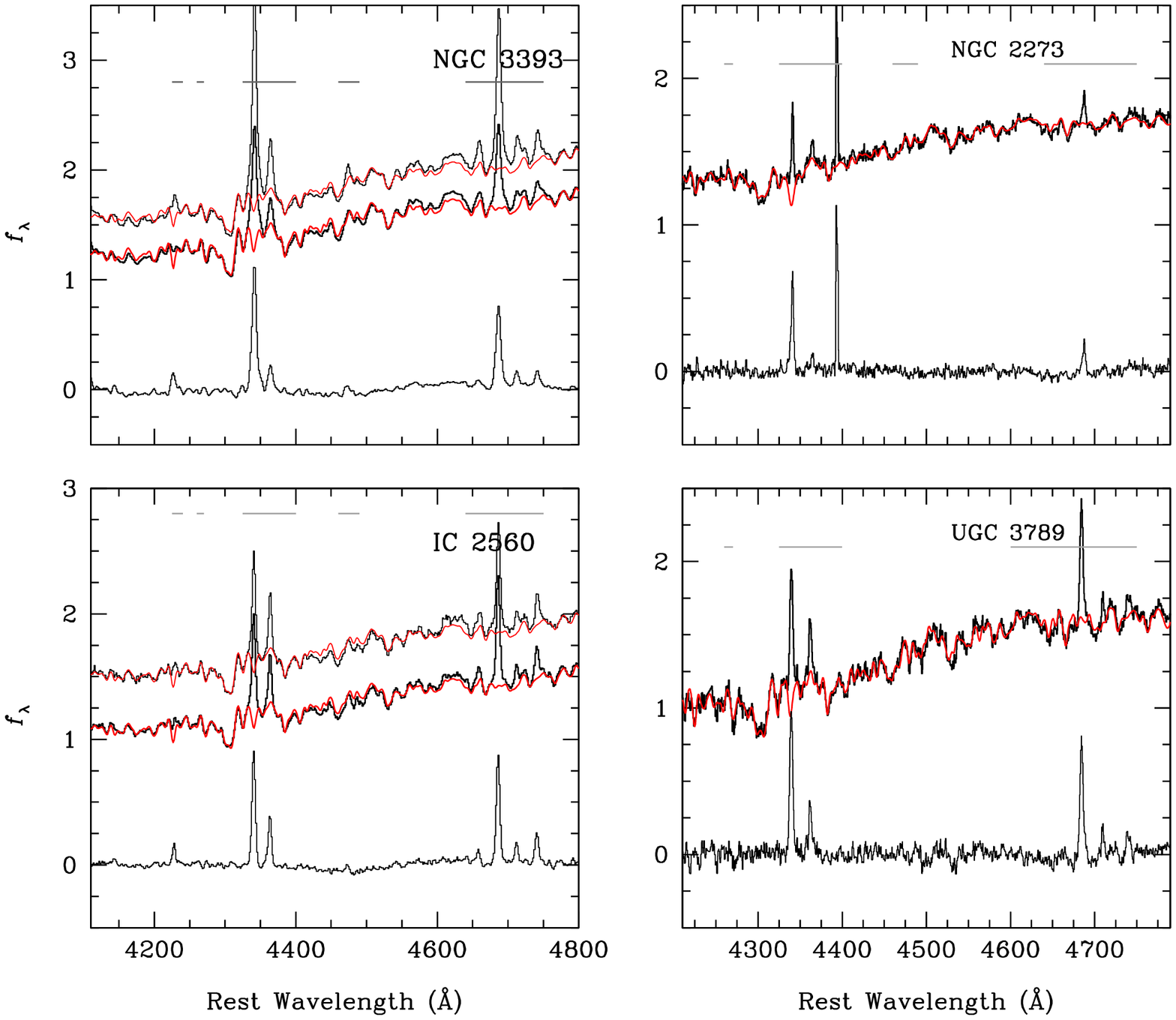,width=0.6\textwidth,keepaspectratio=true,angle=90}
}
\vskip -0mm
\figcaption[]{Example fits to four objects in the sample.  {\it Left}: Stellar
velocity dispersion fits to two B\&C spectra, using the best-fit
weighted sum of an A9, G6, G8, K0, and K3 star.  We plot the B\&C
spectra along the major axis ({\it middle; thick histogram}), the same
along the minor axis ({\it top; thin histogram}), best-fit models
({\it red solid}), and residuals ({\it bottom; thin histogram}), while
regions excluded from the fit are demarcated with grey bars.  
Flux density is normalized to the continuum.  {\it Right}:
Spectra displayed as above, but we have only major axis spectra 
observed with DIS.  Here we used an F2, G6, K1,
and K4 star.
\label{fits}}
\end{figure*}

The primary observations discussed here were obtained using the Dupont
2.5m telescope at Las Campanas Observatory in the South, and the
Apache Point Observatory (APO) 3.5m telescope in the North.  In the
Appendix we describe an auxiliary data set that we use to cross-check
our velocity dispersion measurements.  At the Dupont telescope, we
used the Boller \& Chivens (B\&C) spectrograph with the 600
lines~mm$^{-1}$ grating and the 2\arcsec\ slit, which yielded an
instrumental resolution of $\sigma_r \sim 120$~\kms\ and a wavelength
coverage of $3400-6600$~\AA, covering the 4000\AA\ break, the G-band
at 4304\AA, \hbeta+\oiii$~\lambda \lambda 4959, 5007$, the
\mgb$~\lambda \lambda 5167, 5173, 5184 \AA$ triplet, and (in most
cases) \halpha.  We observed at two slit positions (Table 1) for the
majority of the galaxies, both along the major and minor axis, at the
lowest possible airmass to mitigate differential refraction.  We
observed the Northern targets with APO using the Dual Imaging
Spectrograph (DIS), with a $1\farcs5$ slit, a 1200 lines~mm$^{-1}$
grating, a central wavelength of 4820~\AA, a spectral range of
4250-5400 \AA\ and a dispersion of $\sigma_{\rm instr} \approx
52$~\kms, as measured from the arc lines.  These observations were
performed at the galaxy major axis only.  We also present one calcium
triplet~$\lambda \lambda 8498, 8542, 8662 \AA$ (\cat) measurement,
based on a red setting with a central wavelength of 8540~\AA\
(8000-9100 \AA) and comparable instrumental resolution.

The reduction of both the DIS and B\&C data proceeded in a similar
fashion.  Flat-fielding, bias-subtraction, and wavelength calibration
were performed within {\tt iraf}\footnote{http://iraf.noao.edu/}.  In
the case of the B\&C data, flux calibration and telluric correction
was also performed within {\tt iraf}, using LTT 3218, LTT 3864, LTT
6248, Hilt 600, LTT 3218, Feige 56, and CD 32 as flux calibrator stars
in order to match the airmass and time of observation.  Flux
calibration and telluric correction of the DIS data used IDL routines
as described by \citet{mathesonetal2008}.  In this case we used Feige
34 as the flux calibrator.  In the Appendix we compare the \oiii\
luminosities between two independent data sets and demonstrate that
our flux calibration is good to $\sim 30\%$.

The proper physical aperture used to calibrate the \msigma\ relation
remains a matter of debate
\citep[e.g.,][]{ferraresemerritt2000,tremaineetal2002}, and at present
robust $r_e$ measurements are only available for a small fraction of
the maser sample.  Thus, we have chosen to extract spectra with
apertures ranging from the resolution limit of the observations ($\sim
2\arcsec$) to the S/N limit of the spectra ($\sim 10\arcsec$).
Typically the signal-to-noise (S/N) ratio is optimized with apertures
of $\sim 4\arcsec$.  The resulting dispersion in \sigmastar\ values is 
included in the error budget, and is $\lesssim 10\%$.

\section{Stellar Velocity Dispersions}

We use direct-pixel fitting \citep[e.g.,][]{burbidgeetal1961} to
measure stellar velocity dispersions.  The stellar templates are
broadened and fitted in pixel rather than Fourier space.  
\hskip -0.05in
\psfig{file=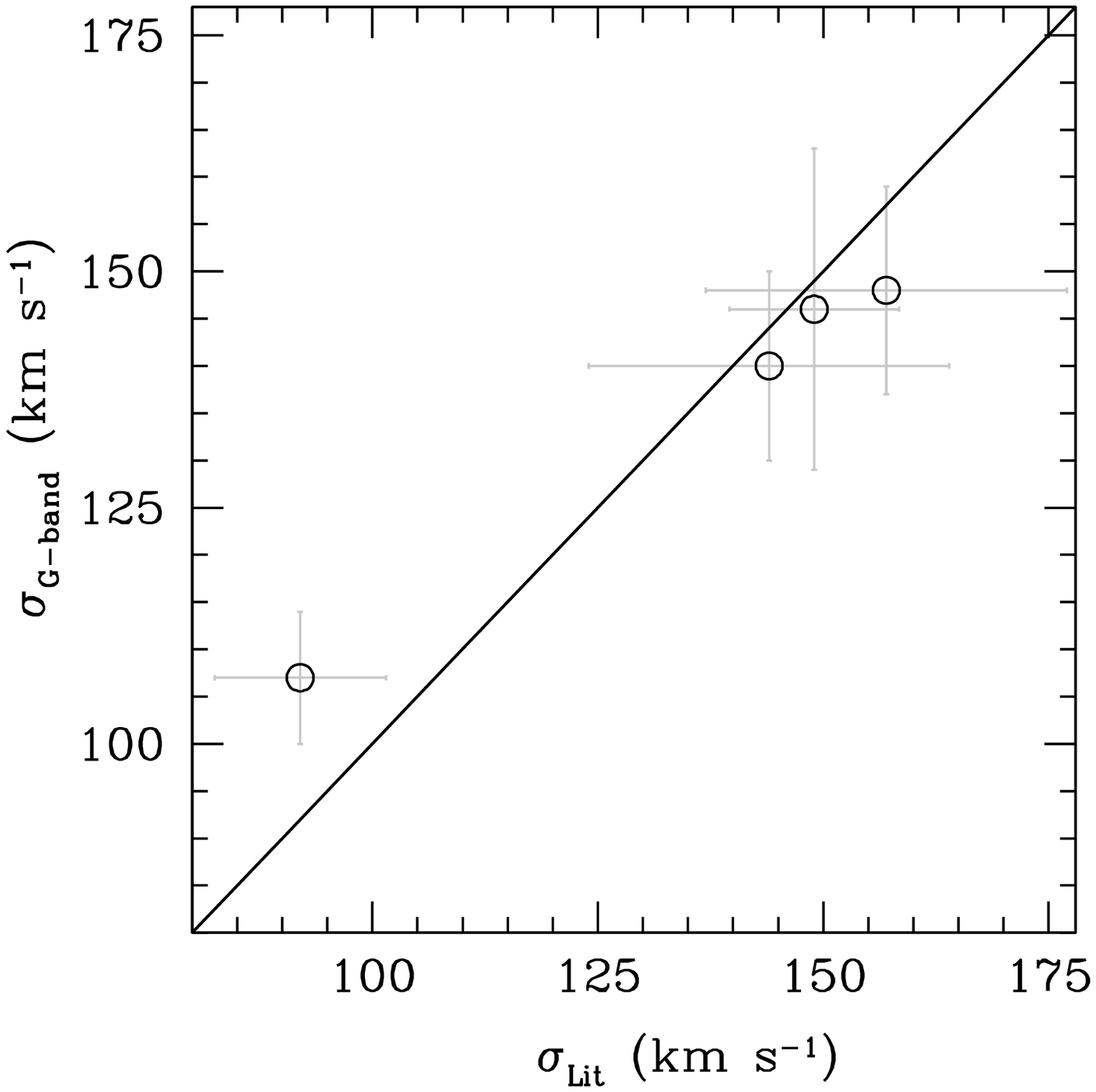,width=0.40\textwidth,keepaspectratio=true,angle=0}
\vskip -0mm
\figcaption[]{
Comparison between our measured velocity dispersions and values from the 
literature (see \S 4.1 for the origin of each dispersion measurement).  Note 
that we have not included the alternate dispersion for NGC 1194 
(\sigmastar$=184 \pm 18$ \kms).
\label{cfsigma}}
\vskip 5mm
\noindent
This method is computationally expensive relative to Fourier methods
\citep[e.g.,~][]{tonrydavis1979,simkin1974,
  sargentetal1977,bender1990}, but by no means prohibitive by modern
standards.  Direct pixel fitting provides many benefits over Fourier
techniques \citep[e.g.,~][]{rixwhite1992, vandermarel1994,
  kelsonetal2000, barthetal2002, bernardietal2003a}.  For one thing,
it is easier to incorporate noise arrays directly.  Of particular
importance to active galaxies, arbitrary regions of the spectrum (such
as narrow emission lines) may be masked in a straightforward manner
without introducing spurious signals into the final results.

The details of the code used here are described in
\citet{greeneho2006sig} and \citet{hoetal2009}.  In short, each
spectrum is modeled as the linear combination of template stars
$T(\lambda)$ that are shifted to zero velocity, broadened by a
Gaussian, $G(\lambda)$, with width $\sigma$ and diluted by a constant
[or power-law if necessary; $C(\lambda)$].  Finally, the entire model
is multiplied by a polynomial [$P(\lambda)$; typically third order] to
account for intrinsic spectral differences, reddening, and errors in
flux calibration:
\begin{equation}
M(\lambda) = P(\lambda) \{ [ T(\lambda) \otimes G(\lambda)] + C(\lambda) \}
\end{equation}
The nonlinear Levenberg-Marquardt algorithm, as implemented by {\tt
  mpfit} in IDL \citep{markwardt2009}, is used to minimize \chisq\ in
pixel space between the galaxy and the model.  We have a library of
template stars observed with the B\&C spectrograph.  For the fitting
we use the following stars: HD 36634 (G6 {\small III}), HD 129505 (G8
{\small IIII}), HD 126571 (K0 {\small III}), HIP 70788 (K3 {\small
  III}), and HD 131903 (A9 {\small V}).  For the APO observations, our
spectral library is much more limited, and we have only one reliable
velocity template star.  Thus, we utilize a large library of high-S/N,
high-spectral resolution ($\sim 26$~\kms) stellar templates from
\citet{valdesetal2004}.  Specifically, we use the following stars: HD
26574 (F2 {\small II-III}), HD 107950 (G6 {\small III}), HD 18322 (K1
{\small III}), and HD 131507 (K4 {\small III}).  In order to translate
the measured dispersion to the true dispersion, we first measure the
instrumental dispersion of the DIS spectra using arc lines (52~\kms),
which yields a quadrature difference of $(52^2-26^2)^{0.5} = 45$~\kms\
between the DIS and Valdes spectra.  We also measure the dispersion of
the DIS velocity template using a Valdes star of the same spectral
type and find $45$~\kms.  The consistency of the two methods is
reassuring. Thus, the final dispersion is simply $\sigma_{\ast} =
(\sigma_{\rm meas}^2 - 45^2)^{0.5}$.  Note that because we use stellar
templates observed with an identical instrumental set-up, we need
apply no corrections to the B\&C spectra.

\begin{figure*}
\vbox{ 
\hskip 15mm
\psfig{file=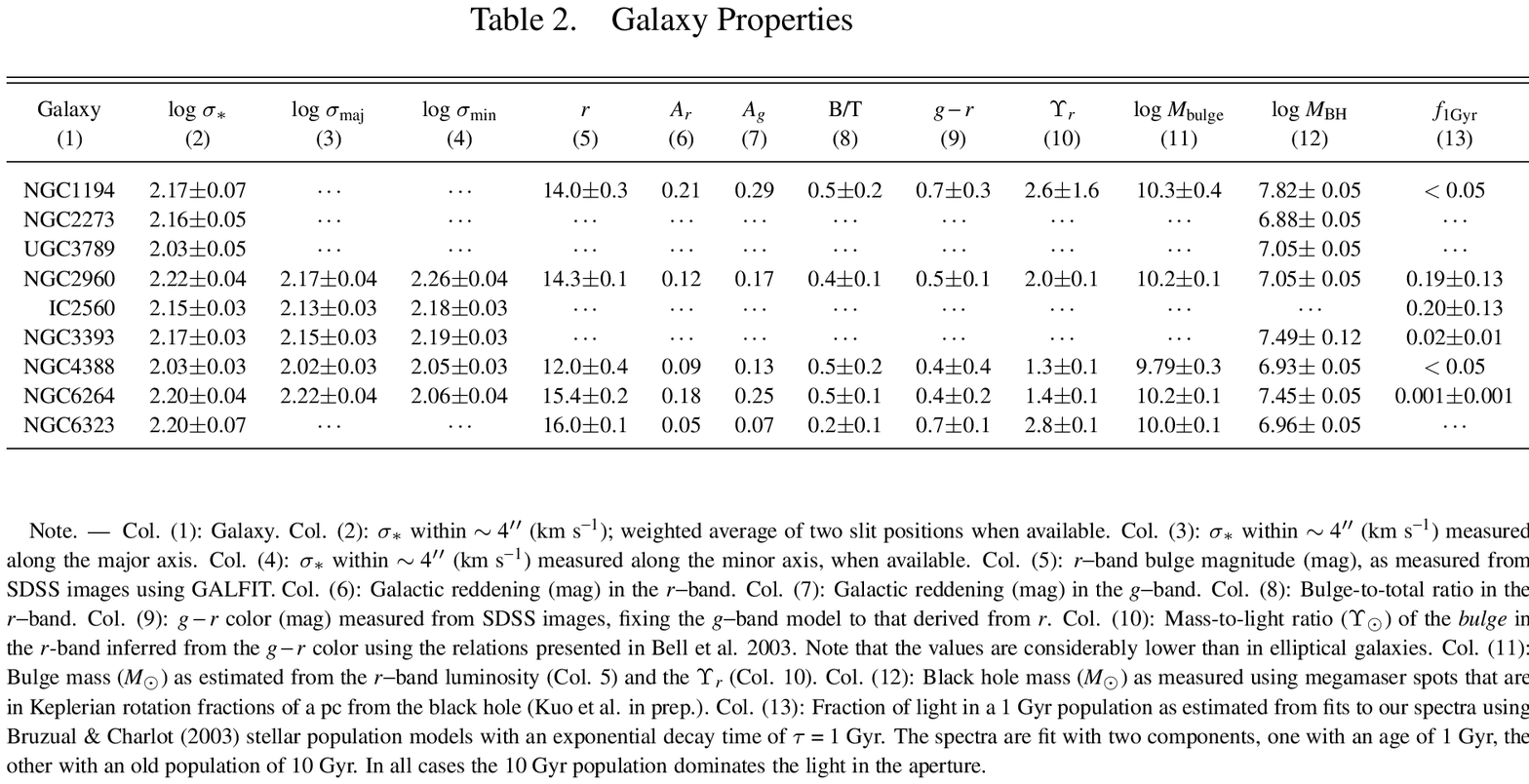,width=0.85\textwidth,keepaspectratio=true,angle=0}
}
\end{figure*}
\vskip 4mm

Given our spectral coverage, the most robust spectral features with
high enough equivalent width (EW) for dispersion measurements are
either the \mgb\ triplet or the G-band.  We avoid the \chk\ features
because they are highly sensitive to spectral type, occur in a
spectral region with a strong continuum discontinuity, and may be
contaminated by interstellar absorption \citep[see review
in][]{greeneho2006sig}.  The \mgb\ features are in many cases
contaminated by emission from [\ion{Fe}{7}]$\lambda 5158$~\AA\ and
[\ion{N}{1}]~$\lambda \lambda 5198, 5200$~\AA.  Furthermore,
differences in the $\alpha$/Fe ratios between the template stars (from
the solar neighborhood) and external galaxies will lead to systematic
errors in \sigmastar\ measurements from this region
\citep{barthetal2002}. For these reasons, we focus on the G-band
region for the velocity dispersion measurements.  We have multiple
slit positions for most of the Dupont spectra.  In Table 2 we present
measurements for each slit position with the $4\arcsec$ aperture, but
we also present the weighted mean of the two, and use that value in
all subsequent work.  Example fits are shown in Figure 2 while the
dispersions are presented in Table 2.

\subsection{Literature Comparison}

There are literature measurements available for a limited number of
targets in our sample.  Direct comparison can be tricky; the
dispersions may be highly aperture dependent
\citep[e.g.,][]{pizzellaetal2004,barbosaetal2006}.  Nevertheless, in
general we find decent agreement with the literature values, as
summarized in Figure 3.  Two targets (NGC 2273, NGC 4388) appear in
the recent atlas of stellar velocity dispersions measured for the
Palomar spectroscopic survey of nearby galaxies
\citep[][]{hoetal1995}, taken with a comparable $2\arcsec \times
4\arcsec$ aperture \citep{hoetal2009}. The two sets of observations
agree within the uncertainties.  For NGC 2273 we find $146 \pm
17$~\kms\ and Ho et al. present $149 \pm 9$~\kms, while for NGC 4388
the values are $104 \pm 10$~\kms\ along the major axis and $92 \pm
10$~\kms\ along the minor axis.  In the case of NGC 4388, note that
our observation is only marginally spectrally resolved.  However, our
SDSS measurement, presented in the Appendix, is $107 \pm 11$, giving
us some confidence in this result. Other literature comparisons for
these two targets are presented in \citet{hoetal2009}.

Two other galaxies (NGC 3393 and IC 2560) were observed by
\citet{cidfernandesetal2004} with a $1\farcs5 \times 0\farcs82$ slit.
Again, the measurements agree within the quoted errors.  In the case
of IC 2560, we find $134 \pm 12$~\kms\ along the major axis compared
to their $144 \pm 20$~\kms, while for NGC 3393 we find $142 \pm
16$~\kms\ along the major axis and $156 \pm 16$~\kms\ along the minor
axis, as compared to $157 \pm 20$~\kms.  NGC 3393 is also presented by
\citet{terlevichetal1990} measured within a $2 \arcsec$ extraction.
They find $184 \pm 18$~\kms, which is marginally consistent with our
minor-axis measurement.

\subsection{Emission Line Spectra and Bolometric Luminosities}

One convenient by-product of our \sigmastar\ measurement technique is
a pure emission-line spectrum that has been corrected for underlying
absorption lines (e.g., \hbeta).  Since the narrow emission lines
(particularly \oiii) represent one of the only methods for estimating
bolometric luminosities for these obscured objects, we describe the
measurements here.  The fitting routines are described in
\citet{greeneho2005o3} and \citet{greeneetal2009}.  Briefly, we fit
the \oiii$~\lambda 4959, 5007$ lines with sums of up to four
Gaussians.  The two lines are restricted to have identical shapes and
the velocity difference and relative strengths are fixed to laboratory
values (3.1 in the latter case).  The \hbeta\ line is fitted
independently with the sum of two Gaussians (which is always
completely sufficient given the typical signal-to-noise ratio, S/N, in
this line).  We also calculate the full-width at half maximum (FWHM)
of the summation of all components in the \oiii\ model.  The velocity
offsets between individual components are small compared to the total
line width, so the FWHM is well-defined.  To bracket systematic
uncertainties, we measure the line width in the spectra from the
4\arcsec and 10\arcsec\ apertures and the error bar represents the
difference in those two widths.

For the majority of the B\&C spectra, as well as the DIS observation
of NGC 6323, we also model the \halpha+\nii\ complex.  Ideally we
would use measurements of the \sii\ line width as a model to deblend
\halpha\ from \nii.  \citep[e.g.,][]{hfs1997broad,greeneho2005o3}.
However, in this case we simply describe each narrow line with the same
combination of Gaussians, fix the relative strengths of the
\nii$~\lambda\lambda 6548, 6583$~\AA\ lines to 1:2.96, and fix the
relative wavelengths of all three lines to their laboratory values.
In the case of NGC 4388, we further include a broad component fixed to
have the same FWHM as reported by \citet{hfs1997broad}, but we are not
sensitive to the presence of weak broad lines in the other objects.
We derive an intrinsic reddening correction for \oiii\ using the
\hbeta\ and \halpha\ fluxes when available, but do not apply it to the
fluxes.  All the line measurements are summarized in Table 3.

We are primarily interested in the \oiii\ luminosities as an isotropic
indicator of the AGN luminosity. Estimating robust bolometric
luminosities in these objects is complicated since megamaser systems
are found preferentially in obscured active galaxies
\citep[e.g.,][]{braatzetal1997}.  Among the systems with circumnuclear
disks, $>75\%$ are Compton thick
\citep[e.g.,~][]{zhangetal2006,greenhilletal2008,zhangetal2010}.
Thus, there are only a few available luminosity diagnostics that are
thought to be isotropic and immune from significant absorption.  One
is the strong and ubiquitous \oiii\ line that has been used both to
find AGNs and to determine their intrinsic luminosities
\citep[e.g.,~][]{kauffmannetal2003agn,zakamskaetal2003} although
concerns about reddening remain \citep[e.g.,][]{mulchaeyetal1994,
  melendezetal2008}.  The shape and equivalent width (EW) of the Fe
K$\alpha$ line at 6.7 keV are also sensitive to the intrinsic
luminosity and inclination of the obscuring material
\citep[e.g.,][]{levensonetal2002}.  In a few cases, detailed modeling
of the Fe K$\alpha$ lines have been performed, which yields an
independent estimate of the bolometric luminosities
\citep{levensonetal2006}.  Finally, \citet{diamondstanicetal2009}
argue for the use of the [\ion{O}{4}]~$\lambda 25.89~\micron$ emission
line as a higher-fidelity tracer of AGN bolometric luminosity
\citep[see also][]{melendezetal2008,rigbyetal2009}.

Using the \oiii\ bolometric correction presented in
\citet{liuetal2009}, we derive bolometric luminosities for the entire
sample using our measurements of \loiii\ (Table 3). Since this
bolometric correction was derived without correction for internal
reddening, we do not deredden either, but tabulate the
proper correction for reference and comparison with the other
indicators.  The estimated Eddington ratios range from
\lledd$=10^{-2}-0.16$.  Including internal reddening would change the
values by 0.6 dex at most.  There are only two objects with estimated
bolometric luminosities from the Fe K$\alpha$ line (excluding IC 2560
for the moment), also shown in Table 3.  There are also three cases
with [\ion{O}{4}] measurements in the literature; we have converted to
bolometric luminosity using the calibration of J.~Rigby et al. in
preparation.  As expected, $L_{\rm bol}$([\ion{O}{4}]) tends to be
higher than $L_{\rm bol}$(\oiii), but the number of objects is too
small to draw definite conclusions about the average magnitude of this
effect, or how directly it is related to our internal reddening
estimates.

\begin{figure*}
\vbox{ 
\hskip 10mm
\psfig{file=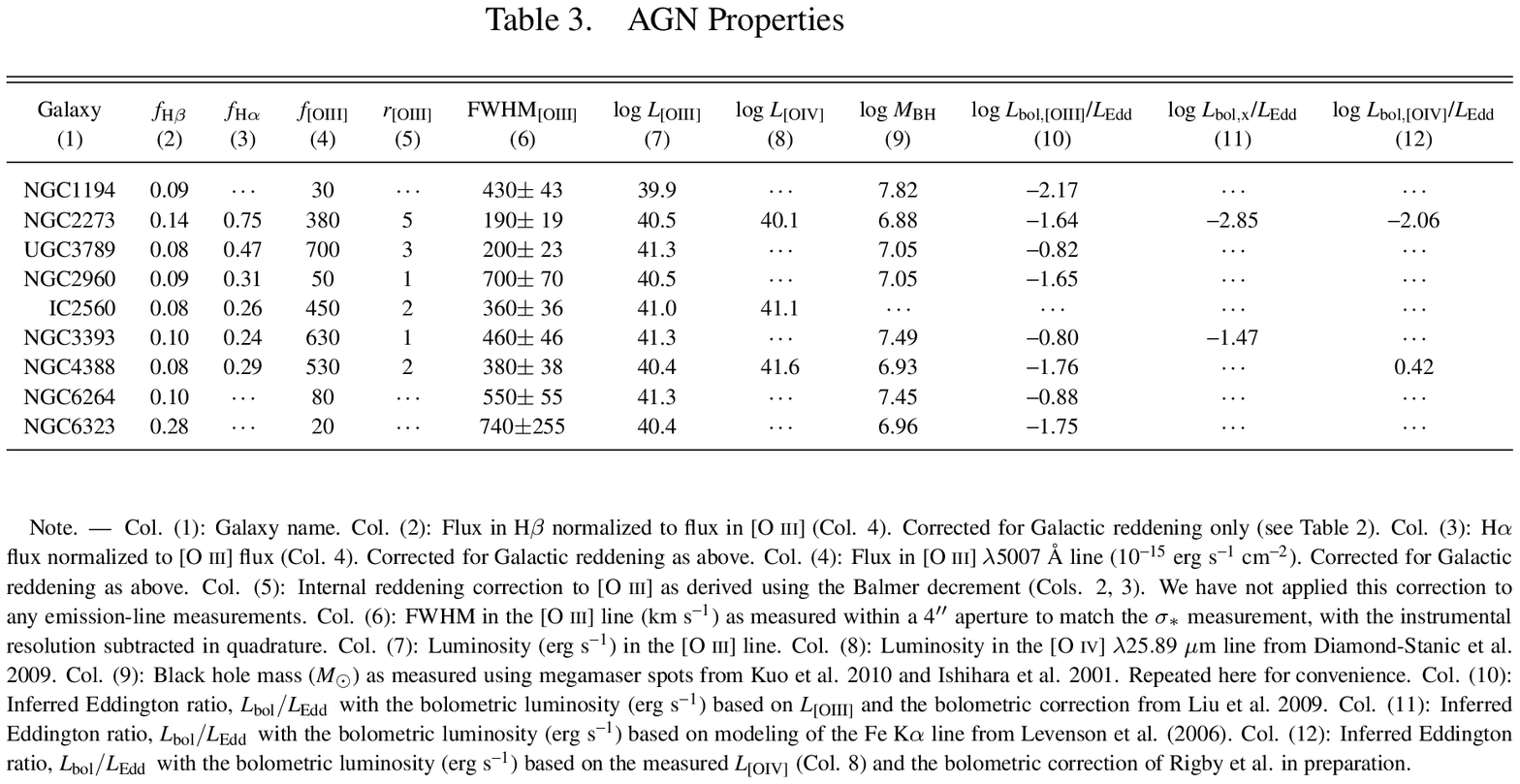,width=0.44\textwidth,keepaspectratio=true,angle=90}
}
\end{figure*}
\vskip 4mm

\subsection{Stellar Populations}

The presence or absence of young stellar populations in the bulge
regions of these galaxies provide important clues as to their growth
and evolution (\S 5.3).  The spectra provide useful constraints on
stellar ages, although the modeling is complicated by the narrow
emission lines that litter the spectrum.  We do not perform full
stellar population synthesis modeling.  Instead, we take a few
representative sets of \citet{bruzualcharlot2003} models and perform a
least-squares fit to the galaxy continua, fixing the stellar velocity
dispersion to that derived above.  We adopt the Padova isochrones
\citep{marigogirardi2007} and a \citet{chabrier2003} initial mass
function.  We take both single-age stellar population models and
exponentially decaying models with a decay time $\tau= 1$~Gyr.  For
simplicity, we use two models with ages of one and 10 Gyr
respectively, although we experiment with models having ages of $0.5$
and 5 Gyr and also one-half of solar metallicity.  We adopt a single
reddening screen for each galaxy, using the reddening law of
\citet{calzettietal2000}.  See example fits in Figure 4.  The fraction
of light in young stars is tabulated in Table 2.

Of the six galaxies for which our spectra include the 4000\AA\ break,
we see that only two show evidence for a significant contribution from
intermediate-age populations (NGC 2960 and IC 2560).  NGC 3393 and NGC
6264 may have a very minor young component, but presumably higher
spatial resolution would be needed to isolate it.  We can compare
these results with the more detailed stellar population fits of
\citet{cidfernandesetal2004}, whose sample also includes IC 2560, NGC
4388, and NGC 3393.  In NGC 3393 we both agree that the stellar
populations are old; Cid Fernandes et al. find that $4\%$ of the
luminosity in this source come from stars with an age $< 25$ Myr.
They find a substantial $33\%$ contribution from young stars in IC
2560, consistent with our findings.  The only ambiguous case is NGC
4388.  They find that $13\%$ of the luminosity comes from this young
component, while we do not find compelling evidence for a young
component.  In fact, NGC 4388 is quite difficult to model, both
because it is extincted and because the emission lines have very high
equivalent widths.  

Those galaxies (NGC 2273, UGC 3789, and NGC 6323) that we observed
with APO do not have the blue coverage needed to constrain the stellar
populations.  In the case of NGC 2273, young stars associated with an
inner ring are well-documented \citep{guetal2003}.  Unfortunately, we
do not have much information for the other two.  The bulge color of
NGC 6323, $g-r=0.7$ (see below), is typical for an early-type spiral
galaxy \citep{fukugitaetal1995}, suggesting that either there is not
much star formation and/or that there is a considerable quantity of
dust.  Overall we find evidence for ongoing or recent star
formation in roughly half of the sample.

\subsection{Uncertainties: Aperture Effects, Rotation, and Bars}

As discussed briefly in \S 2, the velocity dispersion profiles of
late-type spirals can be complicated by kinematically cold, disk-like
components as well as bars or other non-axisymmetries.  As a result,
it is impossible to pick a single, well-motivated aperture size, and
the long-slit data presented here are not deep enough to derive robust
rotation curves.  We can say that the measured dispersions do not
change by more than $10\%$ over the range of radii we probe, and this
spread in values is included in the error budget. Robust $r_e$
measurements are needed to define the optimal radius of extraction for
each galaxy.  Along similar lines, it is possible that the slit
orientation relative to that of the bar might impact \sigmastar\
\citep[e.g.,][]{kormendy1983,emsellemetal2001}.  However, for the
objects with both minor and major axis spectra, we see at most a 40\%
difference in the two measurements.  Ideally, we would like to obtain
integral-field spectroscopy for the objects to settle this ambiguity.

An additional concern is that differing stellar populations and
metallicities lead to bias due to template mismatch.  While our code
includes a mix of spectral types, subtle systematics from region to
region remain \citep[see an example in ][]{hoetal2009}.  When we
calculate errors in \sigmastar, we include the differences between the
nominal dispersions and those measured from the 5100-5400 \AA\ region
[excluding the \mgb\ lines themselves since variations in $\alpha$/Fe
tend to bias the dispersion measurements high
\citep[e.g.,][]{barthetal2002}].

\section{Photometric Decomposition}

Often, spiral galaxies contain bars, ovals, rings, dust lanes,
nuclear spirals, and nuclear star clusters in their centers
\citep[e.g.,][]{carolloetal1997,bokeretal2002,martinietal2003}.
Furthermore, in galaxies with young stellar populations, the
mass-to-light ratio (\ml) may vary significantly both within a given
galaxy and from object to object.  To properly decompose such light
profiles requires both deep imaging on large scales to quantify the
disk component and high resolution nuclear imaging, ideally in a red
band, to reveal the complex nuclear morphology.

We first tried to perform detailed profile decompositions using
Two-Micron All-Sky Survey (2MASS) images \citep{skrutskieetal2006}.
As shown in Figure 5, these images are too shallow to detect
large-scale disks and rings.  Furthermore, the resolution
($2-3\arcsec$) is insufficient to uncover nuclear features.  A
combination of $K-$band observations with \hst\ and from the ground is
needed to fully model the entire galaxy.  As an initial exploration we
have settled on a compromise.  There are five galaxies that fall
within the Sloan Digital Sky Survey footprint.  We have performed
two-dimensional image decomposition on the $gr$ images of these five
galaxies, which allows us to measure a magnitude and a color for their
bulge components.  We then use the relations of \citet{belletal2003}
to convert the $g-r$ color into \mlr\ and thereby stellar mass.

\subsection{\galfit}

We perform image decomposition in two dimensions using the program
\galfit\ \citep{pengetal2002}.  Two-dimensional fitting
allows flexibility in the ellipticities and position angles of
different components that can help break degeneracies in fitting
multi-component galaxies \citep[e.g.,][]{andredakisetal1995,wadadekaretal1999} 
as reviewed in detail in \citet{pengetal2010}.  

\hskip -0.05in
\psfig{file=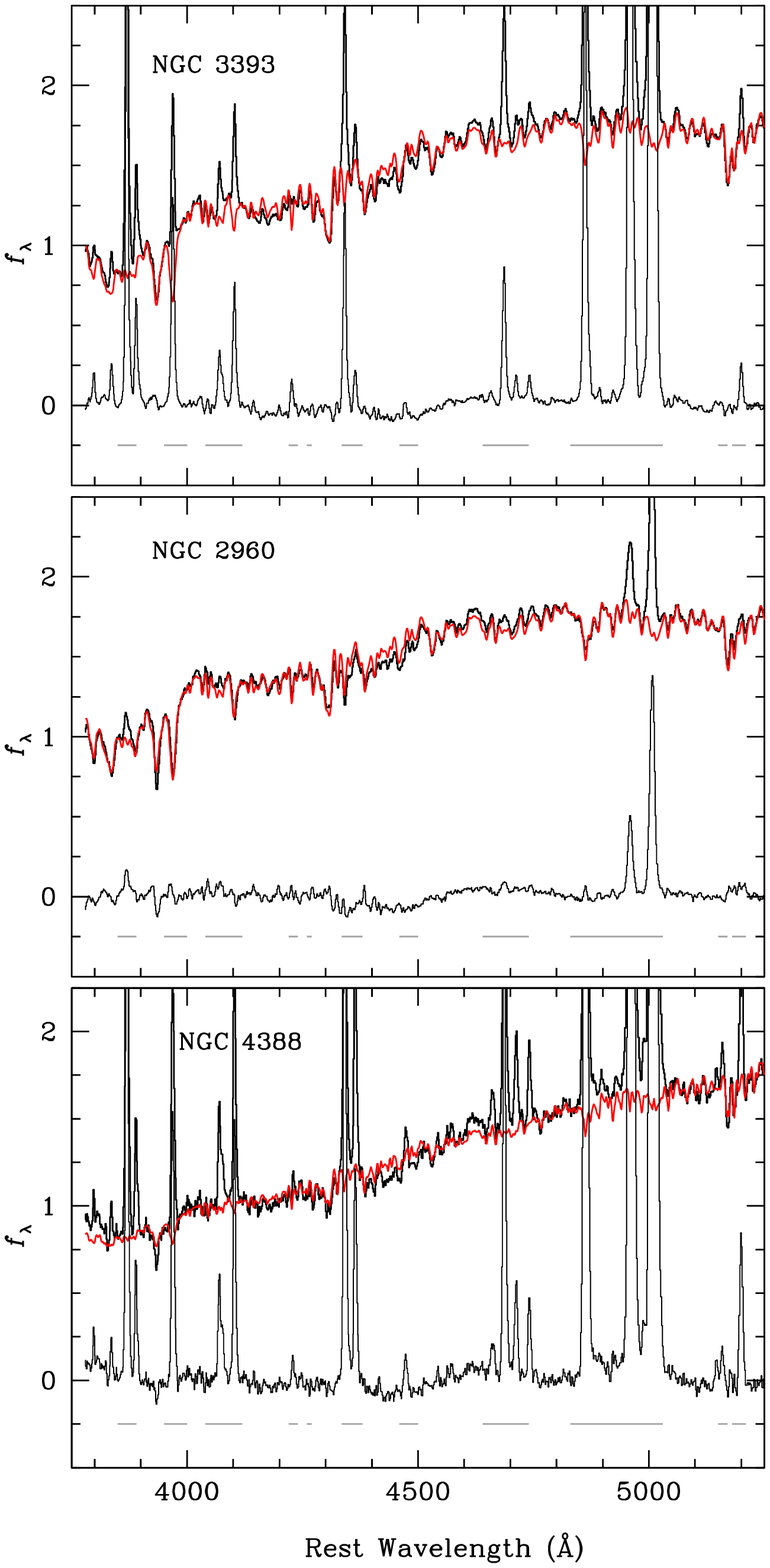,width=0.40\textwidth,keepaspectratio=true,angle=0}
\vskip -0mm
\figcaption[]{Three examples of fits to the entire spectral range using
high-resolution \citet{bruzualcharlot2003} models (see \S 4.3).  
We plot the original spectra ({\it thick black
  histogram}), the best-fit model with \sigmastar\ fixed 
({\it red solid}) and the residuals ({\it thin
  black histogram}).  Flux density is arbitrarily normalized and
masked regions are indicated with grey bars.
\label{cfsigma}}
\vskip 5mm

\galfit\ and related programs create model galaxies as the sum of
ellipsoids with parametrized surface brightness profiles that are
convolved with the point-spread-function (PSF) of the image and then
compared with the data.  Best-fit parameters are those that minimize
the differences between data and model.
\noindent
In general, we model the galaxies with 
\citet{sersic1968} functions:
\begin{equation}
I(r) = I_e~{\rm exp} \left\{ -b_n \left[ \left( \frac{r}{r_e}\right)^{1/n}-1 \right] \right\},
\end{equation}
\noindent
where $r_e$ is the effective (half-light) radius, $I_e$ is the 
intensity at $r_e$, $n$ is the \sers\ index, and $b_n$ is chosen 
such that 
\begin{equation}
\int_0^{\infty}I(r) 2 \pi r dr = 2 \int_0^{r_e} I(r) 2 \pi r dr.
\end{equation}

\hskip -0.1in
\psfig{file=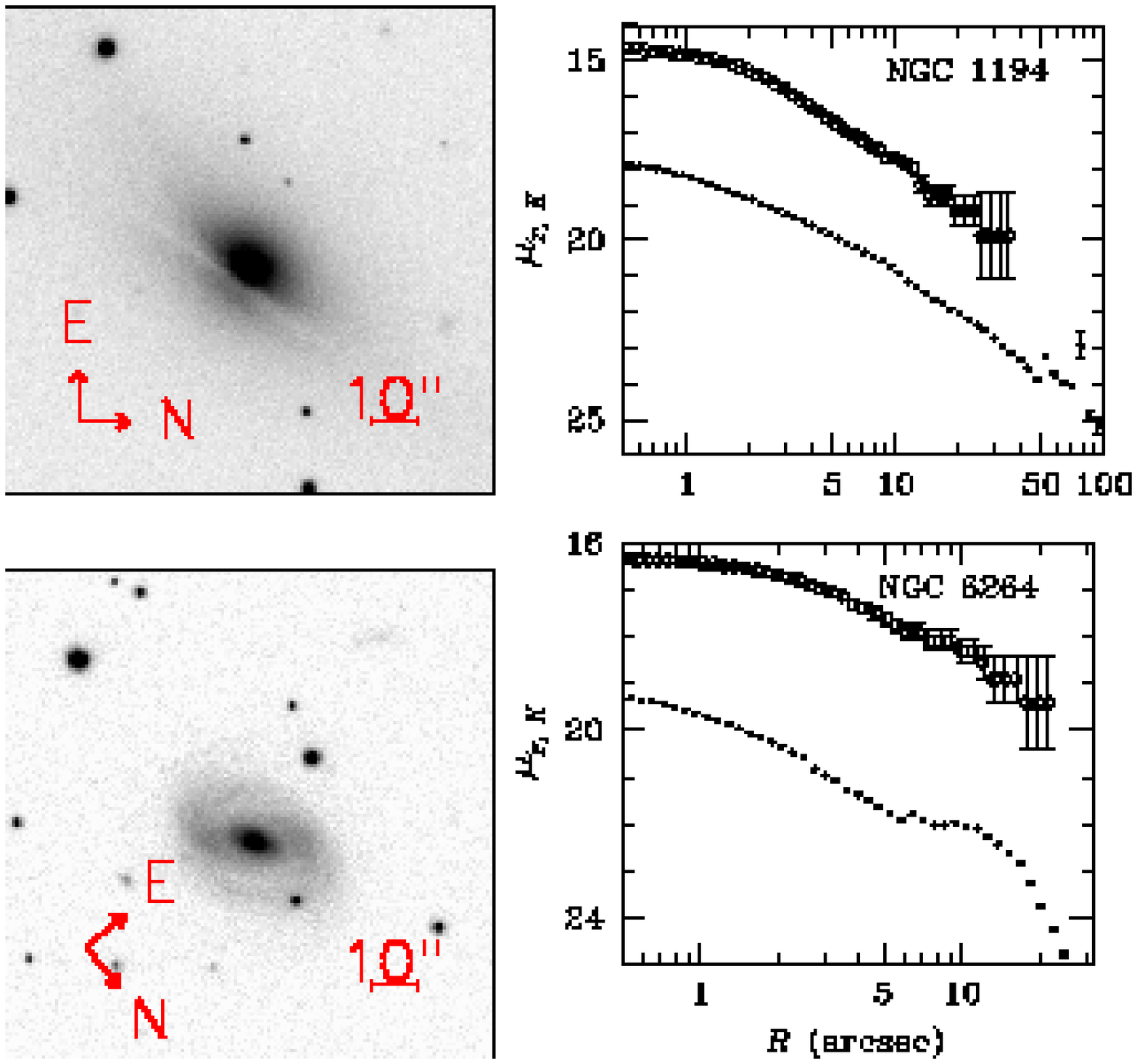,width=0.45\textwidth,keepaspectratio=true,angle=0}
\vskip -0mm \figcaption[]{ 
SDSS $r-$band images ({\it left}) and radial profiles ({\it right}) for the
galaxies NGC 1194 ({\it top}) and NGC 6264 ({\it bottom}).  
A scale of 10\arcsec\ is shown.  The
radial profiles show the $K-$band ({\it open circles}) and SDSS
$r-$band ({\it filled circles}) isophotes, in order to demonstrate
that the optical data are both deeper and have higher resolution.}
\vskip 5mm

An exponential disk has $n=1$ while $n=4$ corresponds to the
well-known \citet{devaucouleurs1948} profile.  We adopt the PSF models
provided by the
SDSS\footnote{http://www.sdss.org/dr7/products/images/read\_psf.html},
and note that our analysis is not very sensitive to the detailed PSF
shape.  Unlike typical two-dimensional modeling, \galfit\ now
incorporates coordinate rotations and Fourier modes (departures from
axisymmetric modeling) that are particularly useful in handling
dust-lanes, bars, and spiral structure \citep{pengetal2010}.

As described above, in addition to the structural measurements we also
need color information to estimate the stellar masses of the galaxies.
We first fit the $r-$band images to extract structural information and
then apply the same model to the $g-$band images, allowing only the
total flux to vary for each component.  In Figure 6 we show the fits
to the $r-$band images.  We caution that the measurements are
preliminary in the absence of high-resolution imaging.  Therefore, we
present only the ``bulge'' luminosities and colors, and reserve
more detailed discussion of the galaxy structure (e.g., \sers\ index)
to future work.  For the moment, we simply note that the bulge colors
provide an additional crude handle on the stellar populations in these
galaxies (Table 2).  In the case of NGC 1194 and NGC 6323, the $g-r$
color is typical of Sa galaxies
\citep{colemanetal1980,fukugitaetal1995}, NGC 2960 has the color of an
Sbc galaxy, and NGC 4388 and NGC 6264 are the bluest, with colors
typical of Scd galaxies.

\begin{figure*}
\vbox{ 
\vskip -0.05truein
\hskip 0.05in
\psfig{file=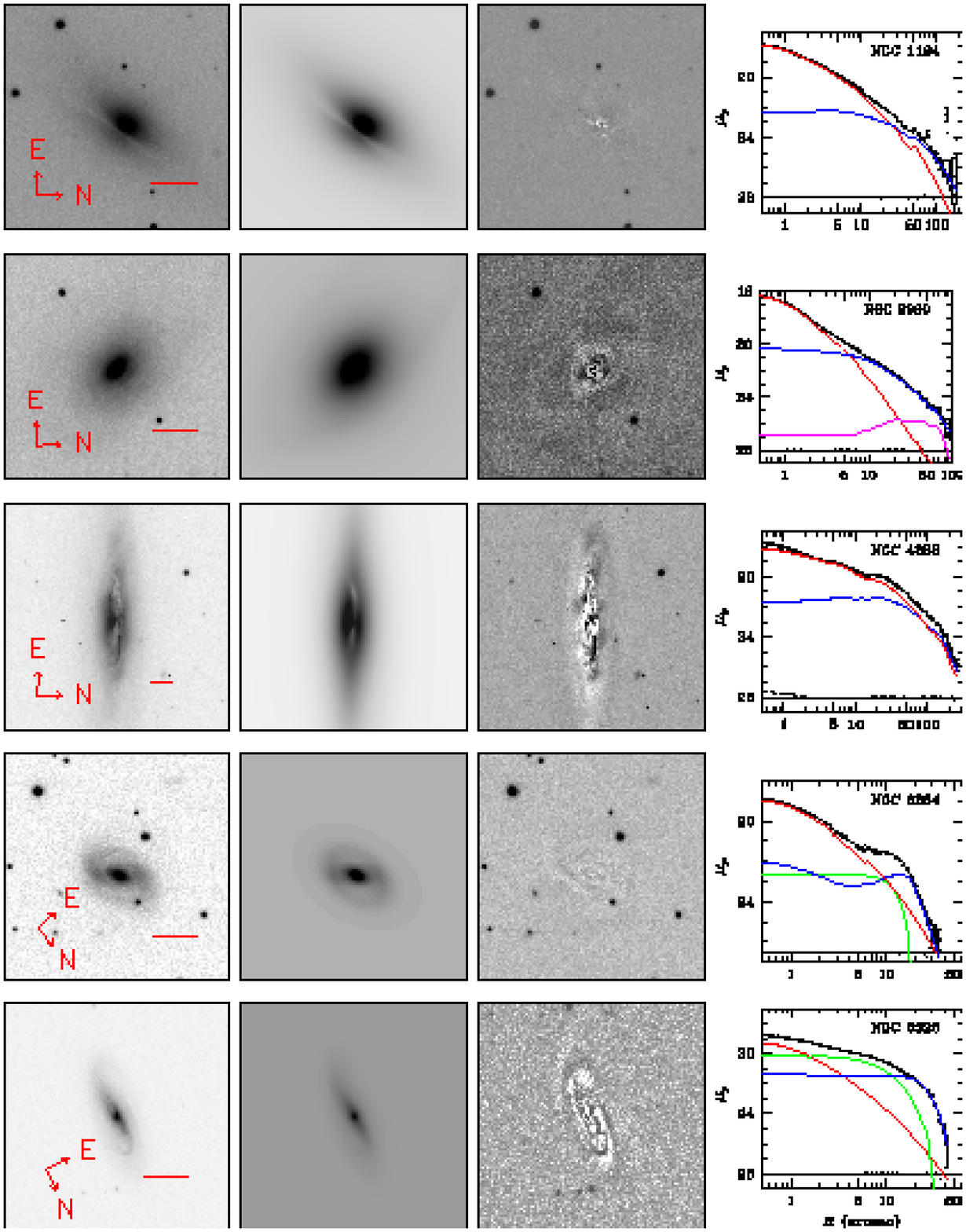,width=0.9\textwidth,keepaspectratio=true,angle=0}
}
\vskip -0mm
\figcaption[]{
Photometric fits to all objects with SDSS images.  We show the $r-$band image
({\it left}) with a scale-bar of 20\arcsec, the best-fit \galfit\
model ({\it middle}), and the residuals ({\it right}), as well as
radial profiles for the data ({\it black circles}), the total model
({\it solid black line}), any bulge ({\it red solid line}), disk
({\it blue solid line}), bar ({\it green solid line}), or outer disk
({\it magenta solid line}).  Residuals are shown beneath in solid
squares. Note that dust-lanes and spiral arms can introduce non-monotonic 
behavior in the radial profiles.
\label{galfits}}
\end{figure*}
\vskip 5mm

\subsection{Uncertainties in Decomposition}

\begin{figure*}
\vbox{ 
\vskip -1mm
\hskip 0.4in
\psfig{file=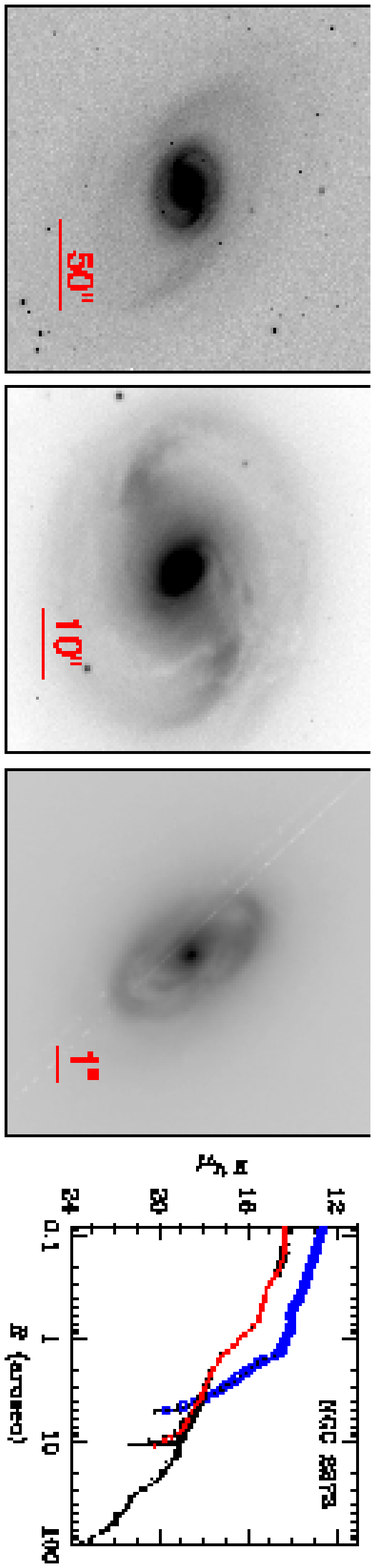,width=0.23\textwidth,keepaspectratio=true,angle=90}
}
\vskip -0mm \figcaption[]{ 
Three images of NGC 2273 at progressively smaller scales.  The 
large-scale $R$-band images are from WIYN and were kindly provided by P. Erwin 
\citep[][; {\it left, middle}]{erwinsparke2003}, while the nuclear 
spiral is a NICMOS/F160W image.  Finally, we show the radial profiles 
for the NICMOS ({\it blue open circles; top}), WFPC2/F606W 
({\it red filled circles}), and $R-$band ({\it crosses}) images.
}
\end{figure*}

Clearly, errors in the PSF shape and sky level add scatter to the
photometric decompositions.  We expect that these uncertainties are
$<0.1$ mag.  Therefore, we are dominated by systematic errors.
More specifically, because of prominent dust lanes in many galaxies,
combined with the limited spatial resolution of the SDSS, we do not
have a strong constraint on the \sers\ index of the bulge, which in
turn contributes large errors to the effective radii as well as 
in the ratio of bulge-to-total luminosity (B/T).

Currently, our error bars represent the spread in derived luminosities
resulting from a range of different models, which give some indication
of the uncertainties due to model assumptions
\citep[e.g.,][]{pengetal2010}.  After running these models, we report
the one with the lowest value of reduced \chisq\ and assign parameter
uncertainties based on the spread across the models. Let us take as an
example the galaxy NGC 1194.  In this case we have fit a model with
the bulge \sers\ index fixed to $n=4$ and the disk fixed to $n=1$.  As
the next level of complexity we have allowed each component to vary
freely, yielding $n=3.9$ for the bulge and $n=2.2$ for the ``disk''.
Finally, we have run two models with different parameterizations of
the dust lane that have $n\approx 2.7, n \approx 1.5$ for the bulge
and disk respectively.  In all cases the fits would be deemed
acceptable, but using a range of models allows us to quantify the
dependence of the bulge parameters on our model assumptions.  In this
case, the resulting bulge magnitude varies by $\sim 0.5$ mag between
the model with $n=4$ and those including dust.  These systematic
uncertainties exceed by far those resulting from sky and PSF
uncertainties, and are the ones we report here.

\subsection{Bulge Classification: Classical or Pseudobulge}

In this subsection we discuss the morphological properties of the
maser sample.  In particular, we address the bulge morphology.  First
consider NGC 1194 (Fig. 5{\it a}).  This galaxy has a massive bulge
component and is classified as an S0.  Both in the radial profile and
the image, it is clear that the bulge in NGC 1194 has smooth isophotes
and a large ratio of bulge-to-total (B/T) luminosities.  Much more
typical of the megamaser disk galaxies is NGC 6264, which has a strong
bar (Fig. 5{\it b}).  Other barred galaxies include NGC 2273, NGC
3393, NGC 4388, and IC 2560.  Roughly speaking, the bar fraction in
the maser sample is consistent with the $\sim 60\%$ seen in the spiral
galaxy population
\citep[e.g.,][]{eskridgeetal2000,menendez-delmestreetal2007}.

Finally, we examine NGC 2273 (Fig. 7), also a barred spiral galaxy.
In addition, NGC 2273 contains two outer rings and a centrally
concentrated bulge component \citep{erwinsparke2003}.  Zooming in with
\hst/WFPC2 and NICMOS reveals that the galaxy center consists of an
inner ring surrounding a nuclear spiral
\citep[e.g.,][]{mulchaeyetal1997,erwinsparke2003}.  This spiral
component is very different in shape, kinematics and stellar
populations from a classical bulge.  The latter is understood to be a
dynamically hot stellar component with an old stellar population,
effectively an elliptical galaxy with a surrounding disk \citep[see
discussion in ][]{kormendykennicutt2004}.  In contrast, the center of
NGC 2273 is a disk with ongoing star formation fueled by a reservoir
of molecular gas \citep[e.g.,][]{petitpaswilson2002} that is likely
fed by the larger-scale bar.  Furthermore, there is a distinct drop in
\sigmastar\ within the inner $5\arcsec$ where the kinematics are
dominated by rotation, presumably because the young bright stars trace
the motion of the gas in which they formed
\citep{barbosaetal2006,falconbarrosoetal2006}.

It has been clear for a long time that the ``bulges'' of spiral
galaxies are often disk-like.  The first clues were kinematic, with
some bulge-like components displaying very high ratios of rotational velocity to
velocity dispersion
\citep[e.g.,][]{binney1978,kormendyillingworth1982,kormendy1993}.
\hst\ imaging reveals disk-like and flattened central isophotes
\citep[e.g.,][]{fathipeletier2003,fisherdrory2008,andredakissanders1994,
  courteauetal1996,macarthuretal2003}, bars, ovals, nuclear spirals
and rings in many bulge-like components
\citep[e.g.,][]{carolloetal1997,martinietal2003} as well as ongoing star
formation \citep[][]{fisheretal2009}.  To distinguish these systems
from ``classical'' bulges, they are referred to as ``pseudobulges''.
It is thought that secular, internal processes (e.g., bars or spiral
arms) are responsible for building pseudobulge mass, in contrast to
the rapid merging processes, i.e. violent relaxation, thought to build
elliptical galaxies and classical bulges.  The literature is reviewed
in detail in \citet{kormendykennicutt2004}.  Throughout, we will use
the term ``bulge'' to refer to any luminous, centrally concentrated
galaxy center, where classical bulges are of the elliptical type and
pseudobulges are essentially disks.

Different authors use different (overlapping) criteria to identify
pseudobulges.  \citet{fisherdrory2008} start with a sample of
morphologically selected pseudobulges, meaning that a nuclear bar,
spiral, or ring is detected within the bulge region \cite[see
also][]{kormendykennicutt2004}.  They show that the distribution of
\sers\ indices in local bulges is bimodal, and suggest that
pseudobulges are galaxies with $n<2$ (for a definition of the \sers\
index, see \S 5.1).  These authors also show that the average B/T of
pseudobulges (0.16) is lower than that of classical bulges (0.4) with
a large spread.  \citet{gadotti2009}, on the other hand, advocates use
of the \citet{kormendy1977} relation as a discriminator, since
pseudobulges tend to have lower central surface brightnesses at a
fixed radius \citep[see also][]{carollo1999,fisherdrory2008}.
Finally, while classical bulges are typified by old stellar
populations, pseudobulges tend to have ongoing star formation
\citep[e.g.,][]{kormendykennicutt2004,droryfisher2007,
  fisheretal2009,gadotti2009}.  Since deriving robust velocity
measurements is beyond the scope of this paper, and in the absence of
more robust structural information, we rely on morphology and stellar
population properties at the present time.

The nearest, well-studied targets in our sample (NGC 4388, and NGC
2273) probably contain pseudobulges.  In the case of NGC 2273 this
classification is based on both the young stellar populations and the
rings and nuclear disk.  NGC 4388 is less certain, but there is clear
evidence for recent star formation and dust.  We suspect that NGC 6264
contains a pseudobulge, given its morphological similarities with NGC
2273 (namely the outer ring and inner bar) and the evidence for young
stars.  The same goes for NGC 3393 and IC 2560, which each contain an
outer ring, a bar, and an inner ring.  On the other hand, NGC 1194,
with both evolved stellar populations and a large bulge, probably
contains a classical bulge.  NGC 2960 has some of the clearest
evidence for ongoing star formation, and so we tentatively put it into
the pseudobulge category.  Finally, we remain agnostic about NGC 6323,
which is one of the most distant targets. Thus, of the nine targets we
consider, at least seven likely contain pseudobulges.

\begin{figure*}
\vbox{ 
\vskip 0.3truein
\hskip 1.3in
\psfig{file=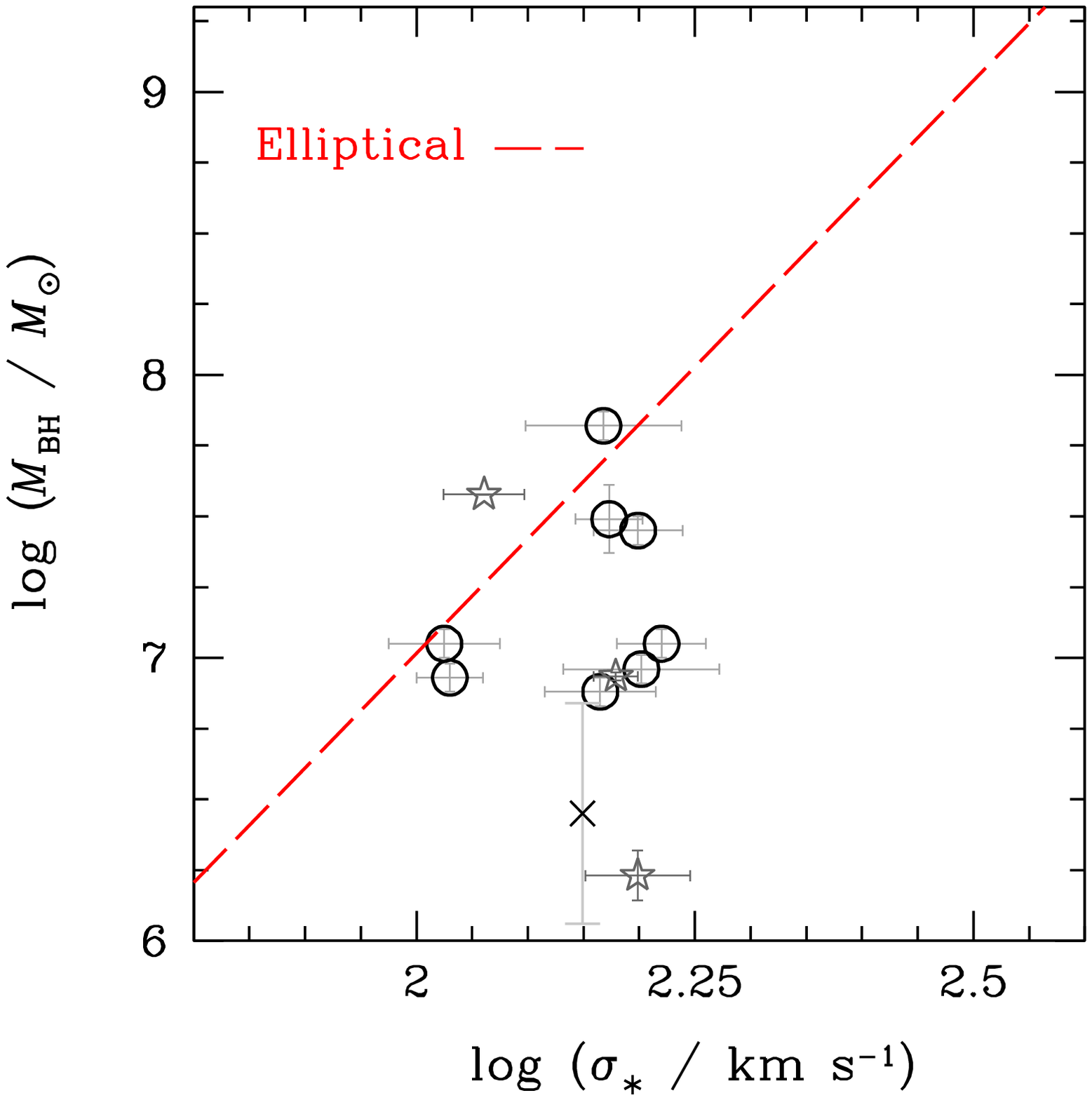,width=0.55\textwidth,keepaspectratio=true,angle=0}
}
\vskip -0mm
\figcaption[]{
The relation between BH mass and bulge velocity dispersion for the
maser galaxies presented here ({\it open circles}) and those from
the literature ({\it grey stars}). IC 2560 is indicated with a cross
and the BH mass error bar is heuristic only.  For reference, we show 
the \msigma\ relation of elliptical galaxies from 
\citet[][ {\it red dashed line}]{gultekinetal2009}.  
The maser galaxies trace a population of low-mass 
systems whose BHs lie below the \msigma\ relation defined by 
elliptical galaxies.  The largest outlier galaxies are (from highest to 
lowest \mbh) NGC 2960, NGC 6323, and NGC 2273.
\label{msigmamaser}}
\end{figure*}

\begin{figure*}
\vbox{ 
\vskip 0.3truein
\hskip 0.5in
\psfig{file=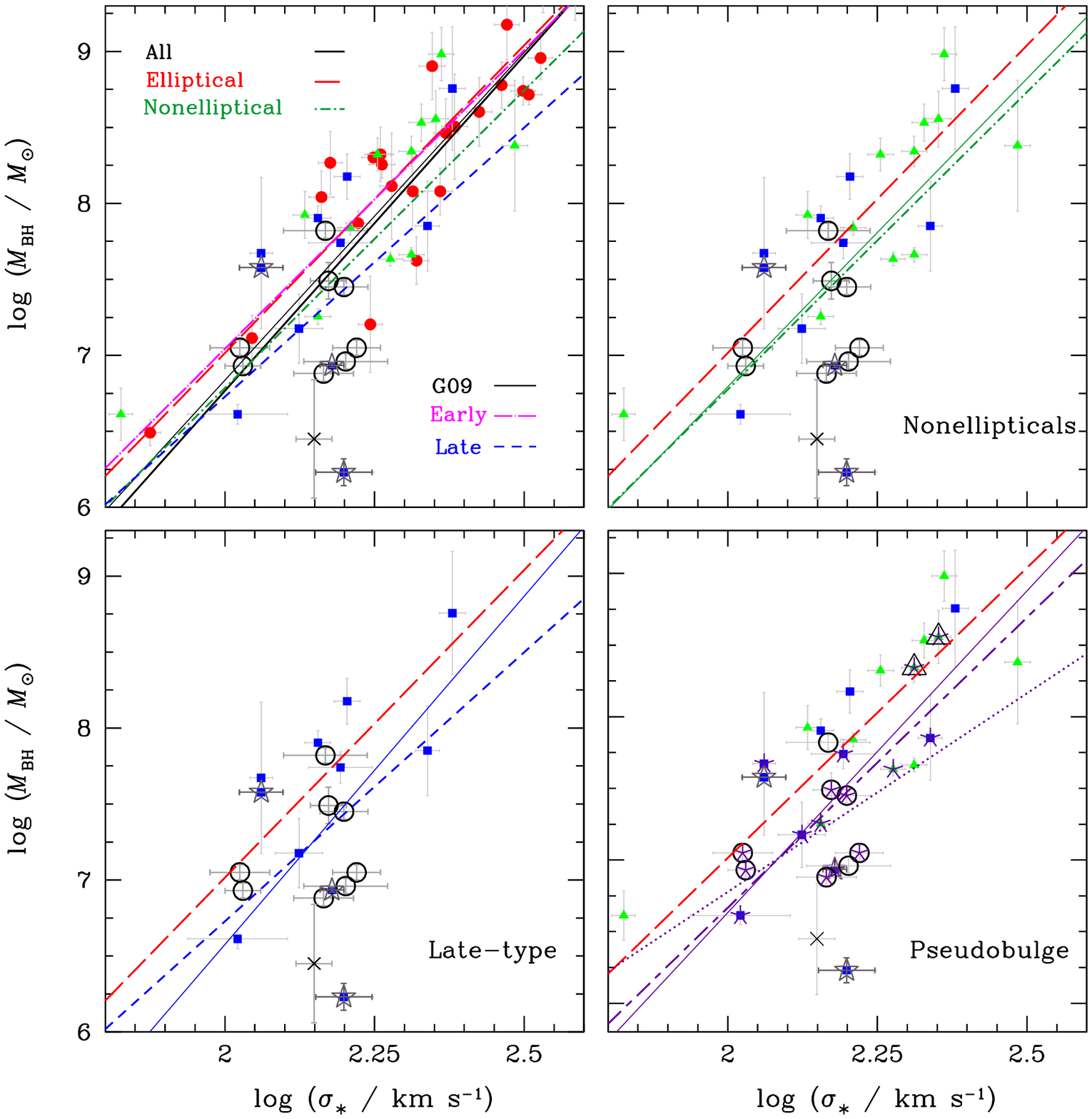,width=0.80\textwidth,keepaspectratio=true,angle=0}
}
\vskip -0mm
\figcaption[]{
The relation between BH mass and bulge velocity dispersion,
including not only the maser sample ({\it open circles}) but also
the sample from \citet{gultekinetal2009}.  Each panel shows fits to
a different subsample (see also \S 6.1). In each panel, literature
maser galaxies are highlighted with open grey stars.  IC 2560 is
indicated with a cross and the BH mass error bar is heuristic only.
For reference, the \msigma\ relation of the elliptical galaxy sample 
 ({\it red dashed line}) is shown in all panels.
({\it Top left}): 
This panel includes all subsamples and all fits, as follows. 
From G{\"u}ltekin et al. we show the elliptical
({\it filled red circles}), S0
({\it green triangles}), and spiral galaxies ({\it blue
squares}).  All of the fits are shown as well, including the
best fit to the entire sample ({\it thick black solid}), the sample of
\citet[][ {\it thin black solid}]{gultekinetal2009}, the elliptical
({\it red long-dashed}) and non-elliptical ({\it green dot-dashed})
subsamples, and to the early (E/S0; {\it magenta long dot-dashed}) and
late-type ({\it blue short dashed}) sub-samples.  For the purpose
of clarity we have omitted upper limits from the figures although
they are included in the fits (see \S 6).
({\it Top right}): 
Here we plot the ``nonelliptical'' sample, consisting of 
S0 ({\it green triangles}) and later-type spiral ({\it blue squares}) 
galaxies.  The original G{\"u}ltekin et al. fit to the nonelliptical 
galaxies ({\it thin green line}) and our fit including the maser galaxies 
({\it green dot-dashed line}) are shown.
({\it Bottom Left}): The ``late-type'' spiral galaxies alone 
(excluding the S0 galaxies; {\it blue squares}). Best fits to the sample 
excluding ({\it thin blue solid}) and including ({\it blue short-dashed}) 
the maser galaxies are shown.
({\it Bottom Right}): The ``pseudobulge'' subsample of the nonelliptical 
sample is highlighted here with purple asterisks.  Note that both S0 and 
later-type spiral galaxies are included, as well as the majority of the 
maser galaxies.  Fits excluding ({\it thin purple solid}) and including 
({\it purple short-long dashed}) the maser galaxies are shown. Furthermore, 
we have performed a second fit ({\it Pseudo2}) excluding NGC 3245 and 
NGC 4342, since they have conflicting morphological designations in the 
literature.  The two galaxies are highlighted with large black triangles, 
and the alternate fit is shown ({\it dotted purple line}).
\label{cfsigma}}
\end{figure*}

\section{Scaling Between \mbh\ and \sigmastar}

In Figure 8 we present the location of the megamaser galaxies in the
\msigma\ plane.  The maser galaxies do not follow the extrapolation of
the \msigma\ relation defined by the elliptical galaxies.  Instead,
they scatter towards smaller BH masses at a given velocity dispersion.
Quantitatively, taking
$\Delta$~\mbh$\equiv$log(\mbh)~$-$~log[$M$(\sigmastar)], where
log[$M$(\sigmastar)] is the expected \mbh\ given \sigmastar, we find
$\langle$\delm$\rangle = 0.24 \pm 0.10$ dex.  There are many hints in
the literature that the \msigma\ relation does not extend to low-mass
and late-type galaxies in a straightforward manner
\citep[e.g.,][]{hu2008,greeneho2008,gadottikauffmann2009}.  However,
the precision BH masses afforded by the maser galaxies make a much
stronger case.  The \msigma\ relation is not universal.  Neither the
shape nor the scatter of the elliptical galaxy \msigma\ relation
provides a good description of the maser galaxies in this plane.

We now add the maser galaxies to the larger sample of local galaxies 
with dynamical BH masses to show that indeed a single, low-scatter 
power-law does not provide an adequate description of all galaxies 
in the \msigma\ plane.  For convenience and to facilitate comparison 
with previous work, we assume a power law for all fits, although 
that form may not provide the best description of the sample as a whole.

\subsection{Subsamples}

We fit not only the full sample, but also divide the galaxies by
morphological type.  For ease of comparison, we adopt the divisions of
\citet{gultekinetal2009}.  In essence, the galaxies are divided into
two primary morphological bins: elliptical and spiral galaxies.  For
completeness we perform fits assuming that S0 galaxies belong to each
subgroup.  We adopt nomenclature from G{\"u}ltekin et al..
``Elliptical'' refers only to the elliptical galaxies, while
``early-type'' refers to elliptical and S0 galaxies.  ``Late-type''
galaxies are spiral galaxies excluding S0s while ``nonelliptical''
includes all spiral galaxies (S0 and later).  All fits are shown in
Table 4 and Figure 9.

In addition to these two divisions, we also consider pseudobulges by
themselves.  Early work suggested that pseudobulges (see \S 5.3) obey
the \msigma\ relation of classical bulges based on a very small sample
of galaxies \citep{kormendygebhardt2001}.  Further study, based on
larger samples, has pointed to differences both in the slope and
scatter of BH-bulge relations in pseudobulges
\citep{hu2008,greeneho2008,grahamli2009,gultekinetal2009,
  gadottikauffmann2009}, but the results are not conclusive,
predominantly because the available number of dynamical BH masses in
pseudobulges is so small.  The maser sample mitigates this problem.
As a fiducial sample, we adopt the pseudobulges tabulated by
G{\"u}ltekin et al..  However, both NGC 3245 and NGC 4342 were
classified as classical bulges by \citet{fisherdrory2008} and
\citet{kormendy2001} respectively.  Thus, we have performed a second
fit (Pseudo2) excluding these two galaxies.  While the slopes differ
between the two fits (although not significantly), the intrinsic
scatter does not change.

\subsection{The Fitting}

To facilitate direct comparison, we adopt the fitting procedure of
G{\"u}ltekin et al..  As is commonly done, we fit the relation: log
(\mbh/\msun)$=\beta + \alpha\,{\rm log (\sigmastar/200\,\kms)}$.  We
use a generalized maximum likelihood analysis that is designed to
incorporate intrinsic scatter in a natural manner.  For simplicity, we
assume a Gaussian distribution in both the measurement errors and the
intrinsic scatter (directly investigated by G{\"u}ltekin et al.).  For
a set of observed points ($M_{i},\sigma_{\ast, i}$), we maximize the
total likelihood:
\begin{equation}
\mathcal{L} = \prod_i l_i (M_{i}, \sigma_{\ast,i}).  
\end{equation}
In the presence of measurement errors, if the likelihood of measuring
a mass $M_{\rm obs}$ for a true mass $M$ is $Q_i(M_{\rm obs}\vert M)
dM_{\rm obs}$, then for a given observation the likelihood is:
\begin{equation}
l_i = \int Q_i (M_i \vert M) P (M \vert \sigma_{\ast, i}) dM
\end{equation}
We assume both $Q$ and $P$ to have a log-normal form.
Upper limits are included, where the mass is taken to be below the 
upper limit with a Gaussian probability given by the measurement 
uncertainty (see details in Appendix A of G{\"u}ltekin
et al.).  Their inclusion does not change the fits significantly.  

Since errors in the independent variable are not naturally included,
we run 2000 Monte Carlo simulations for each fit in which the
dispersion values are drawn from a Gaussian distribution around the
measured value, assuming symmetrical errors in the logarithm.  The
reported error is a quadrature sum of the error from the fit and the
width of the final distribution of fitted values from the Monte Carlo
runs.  As a sanity check, we also employ a \chisq\ fitting technique
adapted from \citet[][]{tremaineetal2002}. In this formalism,
intrinsic scatter is included as an additional term such that the
final reduced \chisq$=1$.  While we only present results from the
maximum likelihood treatment, we note that the fits are consistent
between the two methods in all cases.

The formal fitting quantifies the visual impression of Figure 8.  At
low BH mass (\mbh$\lesssim 10^7$~\msun), galaxies begin to deviate
significantly from the \msigma\ relation of elliptical galaxies.  The
zeropoint is lower by $0.4 \pm 0.2$ dex and the scatter is larger by
$0.2 \pm 0.1$ dex.  It is worth noting that, in agreement with
G{\"u}ltekin et al., the fits to the pseudobulge galaxies are in no
way distinguishable from fits to either the late-type or nonelliptical
subsamples.  Our data demonstrate that the power-law relationship
between \mbh\ and \sigmastar\ is not universal.  Specifically, we find
a population of BHs that apparently have not fully grown to the
present day \msigma\ relation of elliptical galaxies.

\subsection{Potential Caveats}

Our measurement techniques may not yield \sigmastar\ values that are
directly comparable with those in elliptical galaxies.  There is
firstly the question of whether we are contaminated by disk light.
From the SDSS imaging, it is quite clear that the bulge completely
dominates the light within 5\arcsec.  As an additional sanity check,
in Figure 10{\it a} we look for correlations between the galaxy axial
ratio (as proxy for the inclination) and deviations from the \msigma\
relation, $\Delta$~\mbh.  There is no compelling evidence for a
correlation between the two (Kendall's tau value of $\tau=-0.2$ with a
probability $P=0.7$ that no correlation is present), suggesting that
our \sigmastar\ measurements are not strongly biased by rotation in
the central regions.

We still worry about aperture biases in our \sigmastar\ measurements,
because we do not fully sample the radial behavior of \sigmastar.  We
have measured \sigmastar\ within apertures ranging from 2\arcsec\ to
10\arcsec, and the values only vary by $< 10\%$ within this range.  To
reiterate, we are not reporting \sigmastar\ within $r_{e}$ at this
time because we do not wish to fold the (large) uncertainties in the
structural measurements into our dispersion measurements.  In any
case, the radial variations in \sigmastar\ are insignificant on the
scales we can currently measure. On the other hand, as NGC 2273
demonstrates, two-dimensional \sigmastar\ profiles with higher spatial
resolution may well reveal a decline in \sigmastar\ within some galaxy
centers.

Of course, there is the much more serious concern that these galaxies
do contain small classical bulges.  The recent paper by
\citet{nowaketal2010} presents direct evidence that isolating the
classical bulge component in composite systems can bring later-type
galaxies in line with the local elliptical galaxy \msigma\ relation.  We hope
to investigate this concern directly in future work.

\section{The Relation Between BH Mass and Bulge Mass}

Using the subsample of galaxies with SDSS data, we are able to
investigate the correlation between \mbh\ and $M_{\rm bulge}$.  We
first show the \mlb\ relation in Figure 11{\it a}, along with the
best-fit relation from \citet{gultekinetal2009}.  The $V-$band
luminosities are calculated from the $r$-band magnitudes and $g-r$
colors, using a grid of average galaxy colors derived from the
\citet{colemanetal1980} templates.  We caution that the sample is
still small.  Furthermore, we are only spatially resolving the inner
$\sim 500$ pc of these galaxies, while the nuclear spiral in 
NGC 2273 is $< 300$ pc.  At the same time, dust extinction is clearly
significant in all of these systems, and while \galfit\ now includes
dust lanes, the uncertainties are substantial (e.g., B/T can vary by
nearly a factor of two depending on our treatment).

With these caveats in mind, we find that the location and scatter of
the megamaser galaxy bulges analyzed so far is consistent with that of
the early-type inactive sample.  We see a slight tendency for the
megamaser galaxies to be overluminous at a fixed BH mass.  Given that
these are spiral galaxies, with bulge colors that are significantly
bluer than typical elliptical galaxies, the offset may be attributed
to stellar population differences.  Thus, in Figure 11{\it b} we also
show the \mbulge\ relation, where the bulge \mlr\
is inferred from the bulge $g-r$ color (Table 2) using the relation of
\citet{belletal2003}.  We find that these maser galaxies are 
consistent with the fiducial \mbulge\ relation.  On the other hand, we
find no evidence that \mbh\ correlates with total galaxy luminosity or
mass \citep[][ \S 8.2]{jahnkeetal2009,bennertetal2010}.  

\hskip 0.2in
\psfig{file=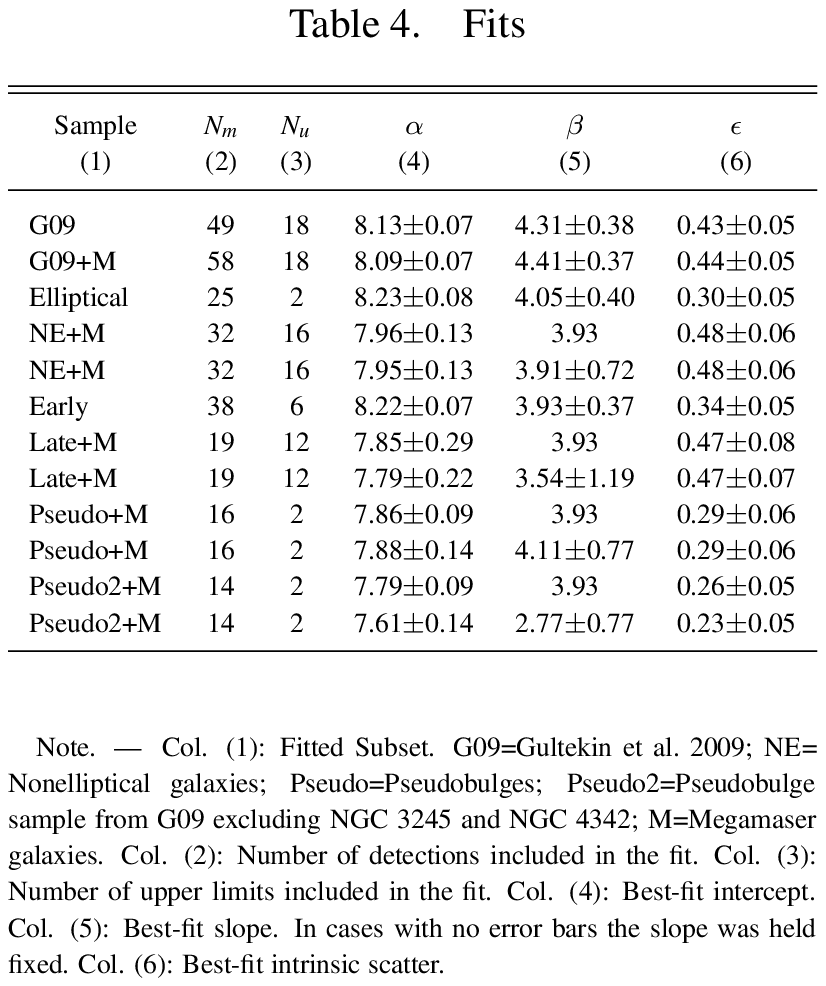,width=0.45\textwidth,keepaspectratio=true,angle=0}
\vskip 4mm

If the \msigma\ relation does not hold for these objects, while
\mbulge\ does, the galaxies ought to be outliers in the
\citet{faberjackson1976} relation, which we briefly investigate now.

\subsection{The Faber-Jackson relation}

In principle, examining the location of the maser galaxy bulges in the
Faber-Jackson relation is an additional tool for evaluating bulge
morphology.  As seen in Figure 12, these five galaxies are consistent
with the Faber-Jackson relation seen in the \citet{gultekinetal2009}
sample.

We have argued that a large fraction of the galaxies in our sample
contain pseudobulges (with the probable exceptions of NGC 1194 and NGC
6323).  The literature on the Faber-Jackson relation in pseudobulges
is limited.  \citet{kormendy1993} find that while the Faber-Jackson
relation of elliptical galaxies defines an upper envelope for the
pseudobulges, they tend to scatter toward low values of \sigmastar\ at
a fixed $B-$band luminosity \citep[see also][]{whitmoreetal1979,
  kormendykennicutt2004}.  The general finding that pseudobulges are
more diffuse at a given luminosity \citep[e.g.,][]{carollo1999} is in
the same sense.  On the other hand, \citet{gadottikauffmann2009}
report the opposite, namely large \sigmastar\ at a fixed stellar mass.
Finally \citet{gandaetal2009} report no difference in the $H-$band
Faber-Jackson relation of late-type spirals, but they are comparing
only to early-type spirals, not elliptical galaxies.  We await a larger 
maser sample to draw definitive conclusions.

\section{Ramifications}

We have demonstrated that there is a population of BHs that scatter
below the \msigma\ relation defined for elliptical galaxies.  In this
section we explore some of the ramifications of a mass-dependent
\msigma\ relation both for our understanding of BH demographics and
for our ability to infer BH masses from \sigmastar\ in low-mass
galaxies.

\begin{figure*}
\vbox{ 
\vskip -0.1truein
\hskip 0.1in
\psfig{file=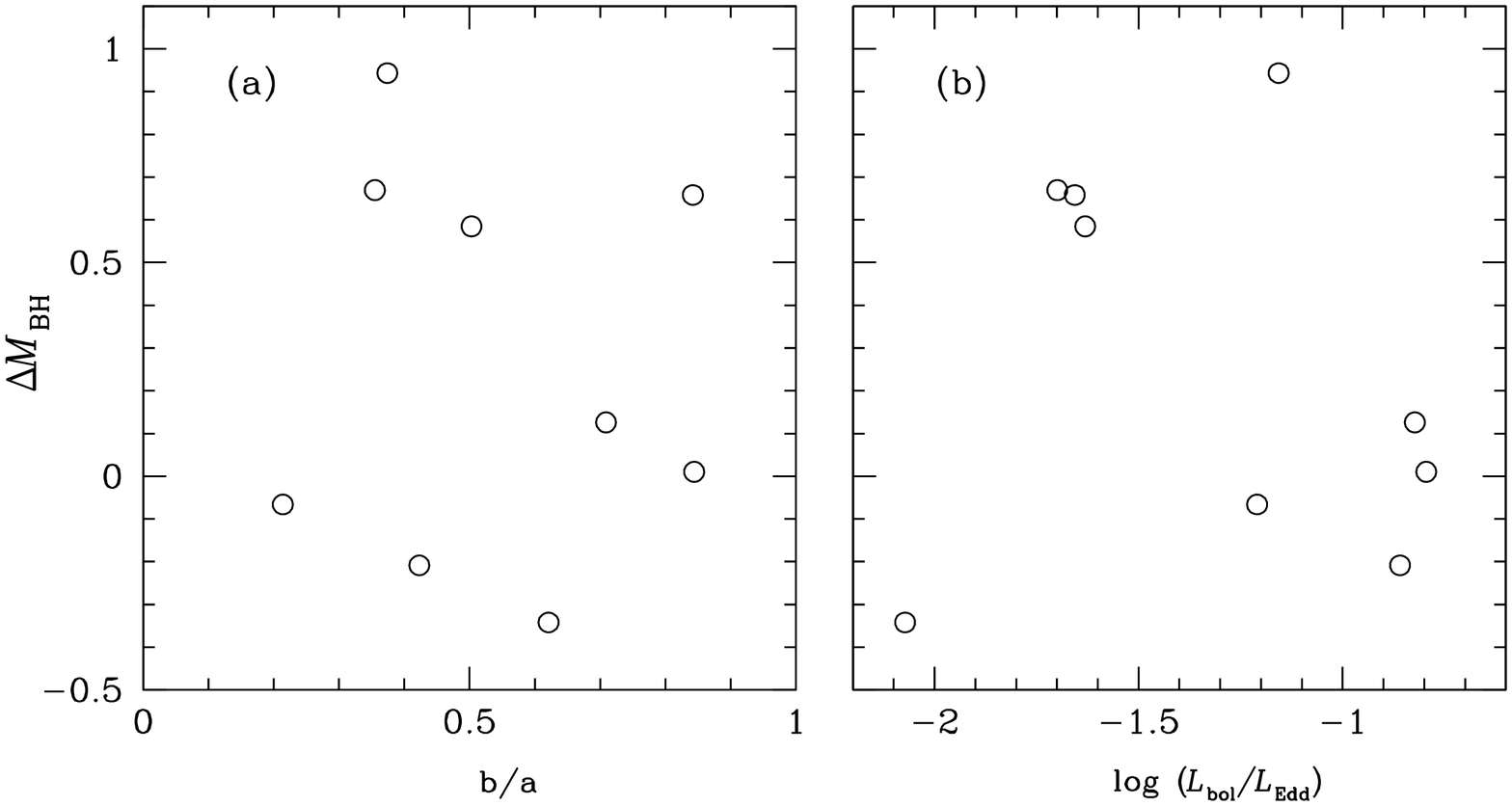,width=0.44\textwidth,keepaspectratio=true,angle=-90}
}
\vskip -0mm
\figcaption[]{
{\bf (a)}: Large-scale disk axial ratio versus deviations from \msigma\ relation;
$\Delta$~\mbh$\equiv$log(\mbh)~$-$~log[$M$(\sigmastar)].  There is
no clear correlation between the two, suggesting that our
\sigmastar\ measurements are not strongly biased by rotation in the
central region. 
{\bf (b)}: We investigate a possible correlation between the 
Eddington ratio \lledd\ and $\Delta$~\mbh.  Again, we see no evidence 
for a significant correlation between the two, suggesting that deviations 
from the \msigma\ relation are not driven by ongoing BH growth. 
\label{incdelm}}
\end{figure*}

\begin{figure*}
\vbox{ 
\vskip -0.1truein
\hskip 0.2in
\psfig{file=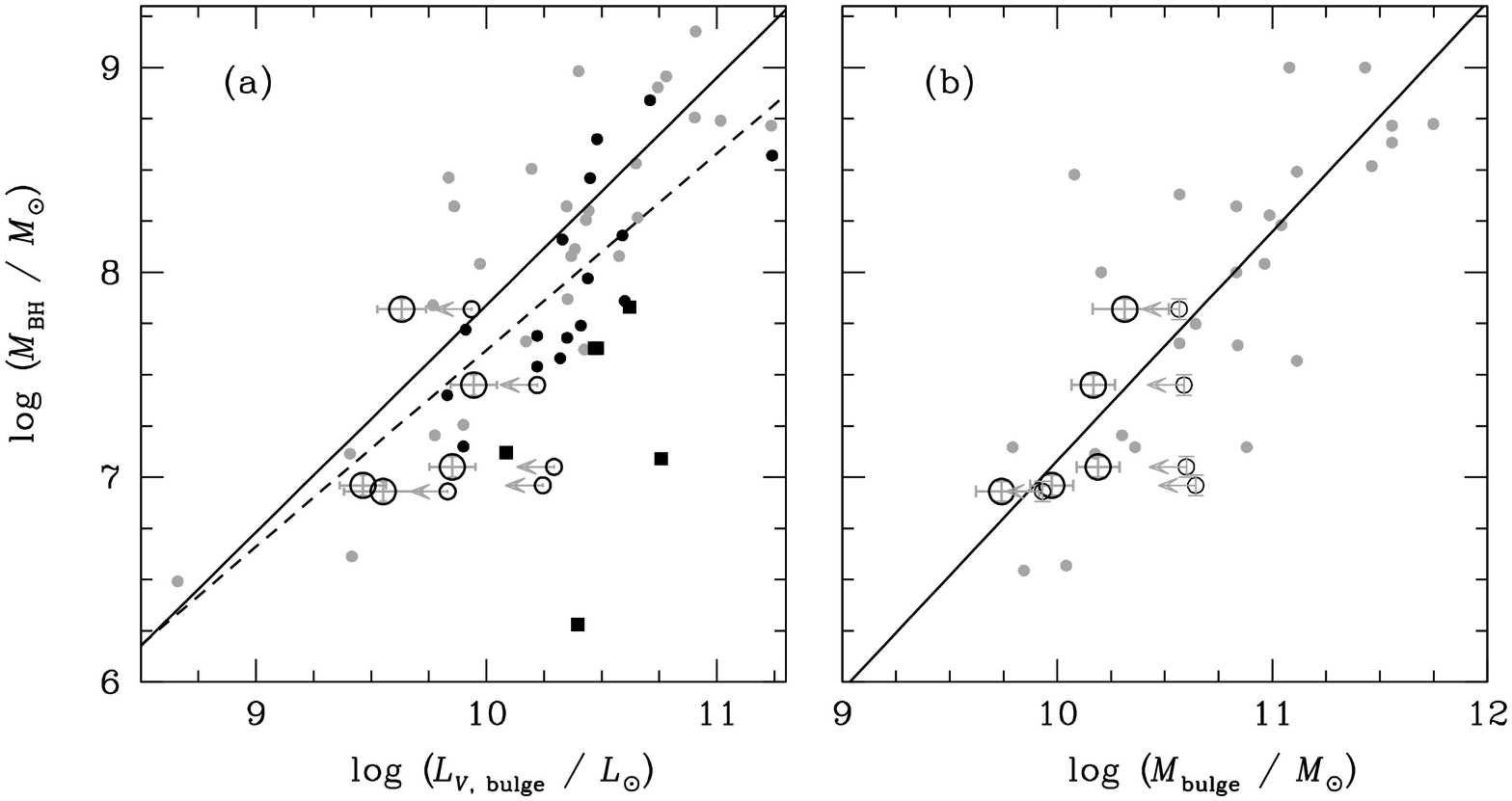,width=0.45\textwidth,keepaspectratio=true,angle=-90}
}
\vskip -0mm
\figcaption[]{({\bf a}): 
The relation between BH mass and bulge luminosity ({\it
  large open circles}) and between BH mass and total galaxy
luminosity ({\it small open circles}) for the megamaser galaxies
with SDSS observations.  Bulge luminosities are derived by detailed
image decomposition of SDSS $r-$band images using \galfit\ (see \S
5, Fig. 7), while the conversion to $V$-band luminosity is
calculated as a function of $g-r$ color using the
\citet{colemanetal1980} galaxy templates.  The solid line shows the
relation of \citet{gultekinetal2009}, while the small grey points
are the objects from their sample in order to demonstrate the
scatter typical in the inactive local samples.  The low-mass galaxy
anchoring the fit to the galaxies from G{\"u}ltekin et al.  is
M32. Note that G{\"u}ltekin et al. fit only E/S0 galaxies.  Filled
black circles are AGNs with reverberation-mapping as presented in
Bennert et al. (2010) here plotted against {\it total} galaxy 
luminosity, while the squares are local AGNs with
reverberation mapping that are not included in the Bennert et
al. (2010) compilation (\S 8.2).  Finally, the dashed line shows the
best-fit relation between \mbh\ and total galaxy luminosity derived
by \citet{bennertetal2010} for AGNs with reverberation-mapping.  In
these local active samples, including the maser galaxies, total
galaxy luminosity does not correlate well with BH mass.
({\bf b}): The relation between BH mass and bulge mass for the same
galaxies as in ({\bf a}).  In this case stellar masses are derived
from the $r-$band magnitude and an estimate of mass-to-light ratio in
the $r-$band (\mlr) derived from the $g-r$ color using the fitting
functions of \citet{belletal2003}.  IC 2560 is indicated with a cross
and heuristic BH mass error bars. The upper limits ({\it small open
  circles}) are total galaxy masses as derived from the $r-$band
magnitude in combination with the \mlr\ for the entire galaxy.  Here
we show for comparison the fit of \citet[][]{haeringrix2004} ({\it solid})
and their sample ({\it small grey points}).  Note that while there is
substantial overlap between the two inactive samples in ({\bf a}) and
({\bf b}), the BH mass measurements are not identical in all cases.
Again we note that BH mass does not correlate with total galaxy mass
in the maser galaxies.
\label{mbhmbulge}}
\end{figure*}

\vbox{ 
\vskip 8mm
\psfig{file=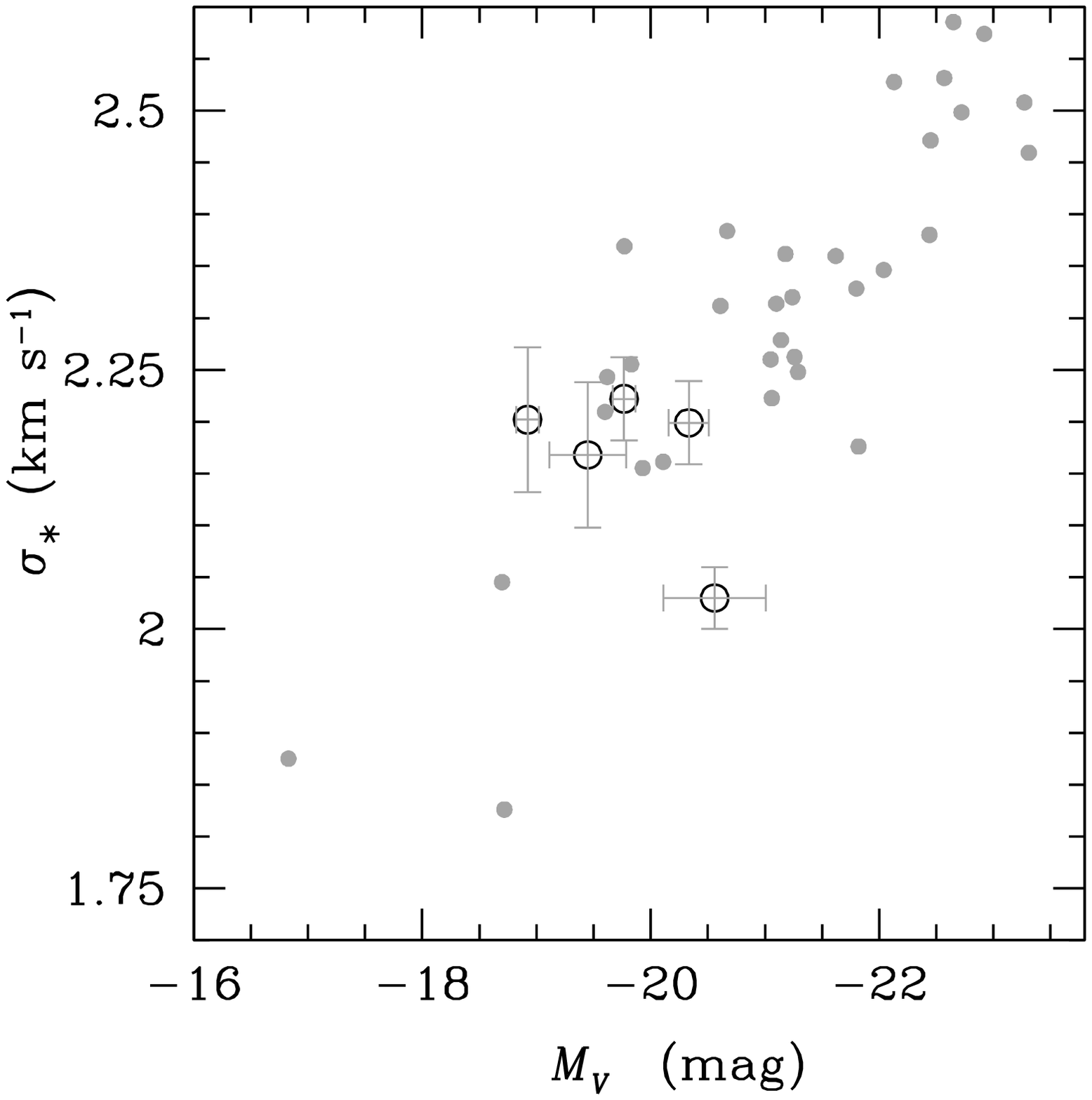,width=0.45\textwidth,keepaspectratio=true,angle=0}
}
\vskip -0mm
\figcaption[]{
The relation between $V-$band magnitude and velocity dispersion of
the five megamaser galaxies with SDSS imaging ({\it large open
  circles}).  On their own, the galaxies do not define any notable
relation between velocity dispersion and magnitude, but they are
consistent with the Faber-Jackson relation seen in the elliptical
and S0 galaxies in the \citet{gultekinetal2009} sample ({\it small
  grey points}).
\label{fjv}}
\vskip 5mm

\subsection{The Local Black Hole Mass Function}

It is common practice in the literature to take the observed
distribution of bulge velocity dispersions or magnitudes and infer the
local BH mass function \cite[e.g.,][]{yutremaine2002,marconietal2004}.
If either the scatter or the shape of these relations change at low
mass, then the shape and amplitude of the inferred BH mass function
for masses $\lesssim 10^7$~\msun\ will also change.  Increased scatter
will tend to broaden the overall distribution, thus increasing the
relative density of objects at high and low mass. The putative offset
and change of slope, in contrast, decrease the overall mass density
for a fixed distribution in \sigmastar.  At present it is unclear
whether the mass functions inferred from bulge luminosity will change.
For this reason, and because \sigmastar\ distributions are not
well-measured in the regime of interest, we do not present a new BH
mass function, but merely note that current estimates of the space
densities of BHs with $10^6 <$\mbh/\msun$<10^7$ are probably
overestimates. This is relevant, for instance, to projected tidal
disruption rates in present and upcoming time-domain surveys
\citep[e.g.,][]{strubbequataert2009}.

\subsection{Normalization of the AGN mass scale}

Currently, the most direct means to estimate BH masses in active
systems comes from reverberation mapping, in which a size scale for
the broad-line region is derived from the lag between continuum
variability and the corresponding variability from the photoionized
broad-line region gas.  Combining the broad-line region size with the
width of an emission line yields a ``virial'' BH mass to within a
scaling factor $f$ that depends on the kinematics and structure of the
broad-line region.  Eventually we hope to measure $f$ in individual
AGNs and we are making good progress \citep[see, e.g.,
][]{bentzetal2008lamp,bentzetal2009lamp,denneyetal2009}.  
\vbox{ 
\vskip 0.2truein
\psfig{file=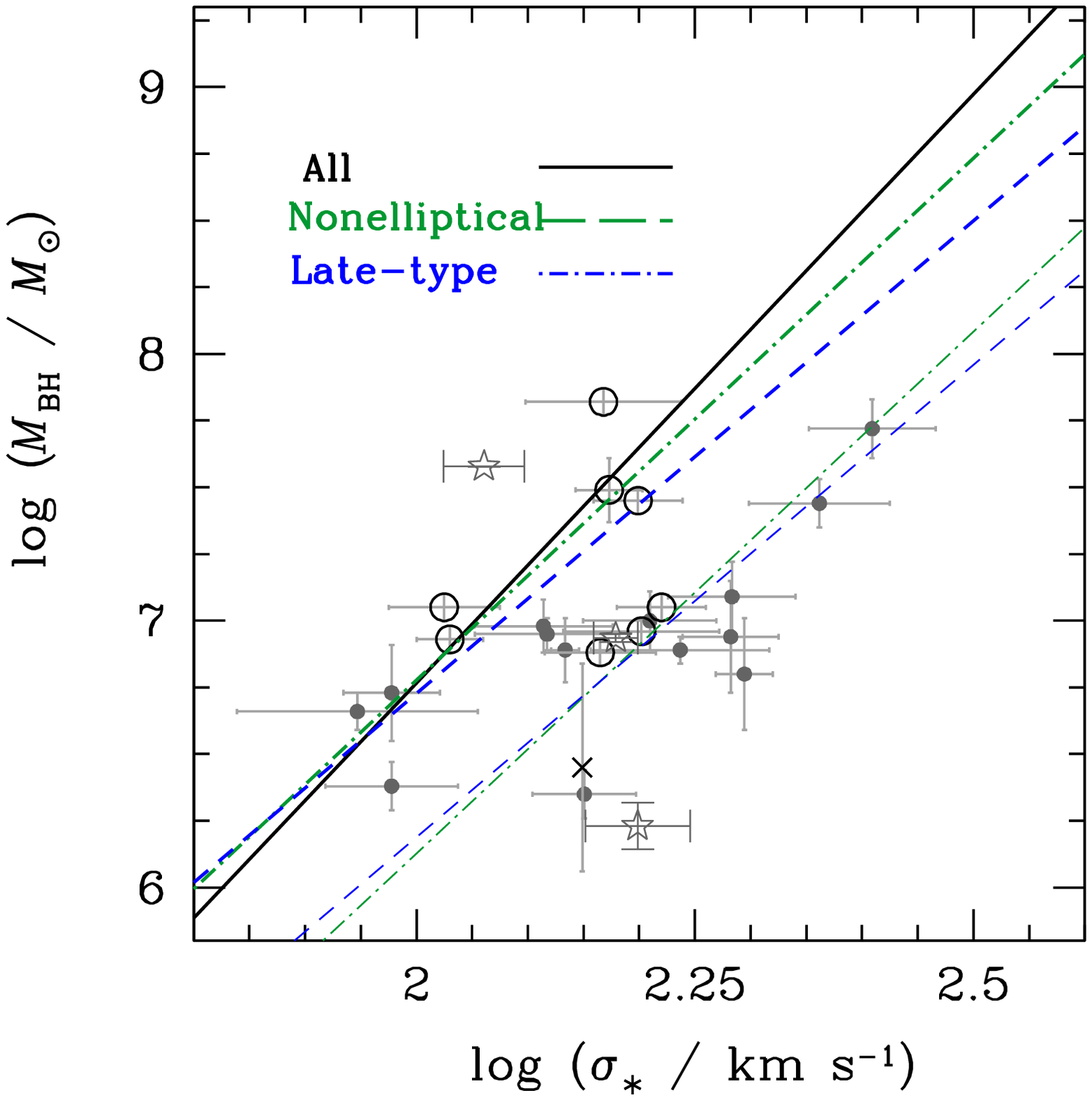,width=0.45\textwidth,keepaspectratio=true,angle=0}
}
\vskip -0mm
\figcaption[]{
The relation between BH mass and bulge velocity
dispersion, now including reverberation-mapped objects ({\it
  grey filled circles}).  We show their virial products (i.e.,
without any assumption about the broad-line region geometry), but we
show the true BH masses for the maser galaxies ({\it large open
  circles}).  IC 2560 is indicated with a cross and the BH mass error 
is heuristic. For reference, the best-fit \msigma\ relations using
all dynamical masses ({\it thick black solid line}), just
nonelliptical galaxies ({\it thick green dot-dashed line}), and just
late-type galaxies ({\it thick blue dashed line}) are shown.
We then determine the best zero-point offset to match the
reverberation-mapped objects with the \msigma\ fits to either the
nonelliptical galaxies ({\it thin green dot-dashed line}) or the
late-type galaxies ({\it thin blue dashed line}).  In order to 
force the active galaxies to obey the local \msigma\ relation for 
late-type galaxies, one needs to multiply the virial products by 
a factor $f \approx 3-4$, differing from the canonical value 
of $f=5.5$ from \citet{onkenetal2004}.
\label{agnmsigma}}
\vskip 5mm
\noindent
For the moment, an average $f$ is derived as the BH mass zeropoint
that brings ensembles of low-redshift, low-mass AGNs into agreement
with the \msigma\ \citep[e..g,][]{nelsonetal2004,onkenetal2004,
  greeneho2006msig,shenetal2008a,wooetal2010} or the \mlb\
\citep{bentzetal2009mbul} relations.  If the \msigma\ relation is
intrinsically different for low-mass or late-type systems, then $f$
must change as well, since the majority of the systems used to derive
$f$ have low masses and late types \citep{bentzetal2009mbul}.

In order to illustrate the sense of the shift, we derive $f$ here
based on our new \msigma\ relation for either the nonelliptical
(including S0 galaxies) or the late-type (excluding S0) galaxies.
Note that we are ignoring the large uncertainties in the measured
slope and zeropoint, simply as illustration. Our procedure is very
simple.  We fix the slope of the relation as derived from one of the
fits above, and then perform a least-squares fit for the zeropoint of
the active galaxies using the unscaled virial products (e.g.,
$\upsilon^2$\re/$G$) as tabulated by \citet{bentzetal2009rl}.

Starting with the \msigma\ relations (Table 4) we derive $f=4.4, 3.4$
using the nonelliptical and late-type fit respectively.  The most recent 
calibration of $f$, from Woo et al., is $f=5.2$.  Thus, we infer that $f$
may be a factor of $\sim 1.5$ lower than the nominal value used in the
literature (Fig. 13).  In certain contexts a difference of this size
matters.  Most notably, studies looking at potential evolution in the
\msigma\ or \mbulge\ relations with cosmic time uniformly adopt the
$f$ value from Onken et al. \citep{treuetal2007,wooetal2008}.
However, the \citet{treuetal2007} sample is clearly composed
predominantly of late-type galaxies.  If they were to adopt a lower
value of $f$, then the evidence for redshift evolution, which they
report to be a factor of $2-5$ at $z=0.4$, could disappear
\citep{wooetal2008}.  
We note also that the mild offset that is apparent in the \msigma\
relation of the lowest-mass AGNs
\citep[e.g.,][]{barthetal2005,greeneho2006msig} would also disappear
with this revised scaling. On the other hand, scaling relations at
these low masses are particularly complicated, given that we have no
direct \mbh\ measurements \citep{greeneho2008}.  \citet{bentzetal2009mbul}
find a very well-defined relation between BH mass and bulge luminosity
for active galaxies with reverberation mapping.  With the proper
imaging in hand, we will attempt the same exercise using the \mbulge\
relation of the maser galaxies.

We wish to make one more point related to active galaxies in
particular.  Two recent papers have found that {\it total}, rather
than bulge, luminosity correlates more tightly with \mbh\ in active
systems and that the relation does not evolve with cosmic time out to
$z \approx 1$ \citep{jahnkeetal2009,bennertetal2010}.  The favored
explanation is that secular processes are growing the bulge components
from the galaxy disks such that the correlation with total galaxy
luminosity is preserved.  In Figure 11{\it a} we show the Bennert et
al. relation between {\it total} galaxy luminosity and BH mass
measured for local reverberation-mapped sources ({\it dashed
  line}). In addition, we include a number of nearby
reverberation-mapped galaxies that both Bennert and
\citet{bentzetal2009mbul} exclude from their samples (NGC 3227, NGC
3516, NGC 4051, NGC 4151, NGC 5548, NGC 7469)\footnotetext{In order to
  calculate a galaxy luminosity without AGN contamination, we remove
  the nuclear luminosity as measured by \citet{hopeng2001} and convert
  to a $V-$band magnitude using $B-V=0.8 \pm 0.2$ mag
  \citep{fukugitaetal1995}}.

Unlike the Bennert objects, neither the NGC galaxies nor the maser
galaxies obey a correlation between total galaxy luminosity and
\mbh. Indeed, the lack of correlation with total galaxy mass in
inactive samples was an original motivation to focus on bulge
properties \citep[e.g.,][, Kormendy, J. et al. in
preparation]{kormendyrichstone1995}.  What we do not understand at the
moment is why, for certain subsets of the reverberation-mapped and
higher-redshift AGN samples, total galaxy luminosity appears to
provide a tighter correlation with \mbh.  We are tempted to blame
uncertainties in the zeropoint of the AGN masses (see above).
However, we do not yet have a full explanation for this apparent
discrepancy.

\subsection{Bias in the Active Sample}

Ideally, we would like to use our results to draw more general
conclusions about the population of BHs with \mbh$\approx 10^7$~\msun\
that live predominantly in spiral galaxies.  However, we must also
consider the possibility that the maser sources represent a biased
sample. First of all, the selection process may be biased by the
presence of nuclear activity.  Were we dealing with distant quasars,
we might be selecting on BH mass, while in local inactive systems we
select on galaxy properties.  However, these AGNs are heavily
obscured, and at nearly every wavelength stellar light dominates the
galaxy luminosity on large spatial scales.  Thus, the masers were
selected first by virtue of their galaxy luminosity and then based on
emission-line ratios and we do not expect a large selection bias based 
on nuclear activity.

A more serious concern is that we do not see a representative snapshot
of the galaxy population when focusing on accreting BHs.  Active BHs
have not necessarily reached their final masses, and thus may be
growing towards the inactive \msigma\ relation
\citep[e.g.,][]{grupemathur2004,hoetal2008b,kimetal2008b}.  The BHs
presented here are radiating at $\lesssim 10\%$ of their Eddington
limits, and thus would need $> 1$ Gyr to grow onto the inactive
relation. This is somewhat long compared to the $\sim 10^7$~year
lifetimes typically estimated for quasars
\citep[e.g.,][]{martiniweinberg2001}.  Typical growth times for
pseudobulges, in contrast, are probably many Gyr
\citep[e.g.,][]{fisheretal2009}.  Therefore, the notion that these
particular BHs knew to accrete material at this moment in order to
move towards the inactive relation seems somewhat contrived.  In
addition, we plot \delm\ against Eddington ratio in Figure 10{\it b}.
There is no correlation between these two quantities (Kendall's
$\tau=-0.2$, probability of no correlation is $P=0.7$).  We see no
evidence that the maser galaxies provide a biased tracer of BH-bulge
scaling relations.

\section{Summary}

We consider the host galaxy properties of a new sample of active
galaxies with megamaser disks. Fitting of the maser spots yields BH
masses with $< 15\%$ precision (C.~Y.~Kuo et al.  in preparation).
BH mass measurements in the maser galaxies more than double the sample
of galaxies with dynamical BH masses of \mbh$=8 \times 10^6 - 7
\times 10^7$~\msun, as well as increasing significantly the number of
spiral galaxies with dynamical BH mass measurements.  Our primary
results are as follows:
\begin{enumerate}
\item
There is a population of spiral galaxies, probed by the maser measurements, 
with BHs that systematically lie below the \msigma\ relation of massive 
elliptical galaxies.  As a result, the \msigma\ relation fits to later-type 
and lower-mass galaxies display both a larger scatter and lower zeropoint 
than fits to elliptical galaxies alone (\S 6).  There is no universal 
power-law \msigma\ relation.

\item 
From our limited data set, the galaxies appear to obey scalings
between \mbh\ and bulge luminosity or mass.  On the other hand \mbh\
does not correlate strongly with total galaxy luminosity for the
maser galaxies (\S 7).

\item
To rule out definitively the presence of small classical bulges
buried in the bars, dust, and young stellar populations found at the
centers of these galaxies, we require spatially resolved kinematics
and high-resolution imaging, which will fully characterize the
complicated bulge regions of these galaxies.

\item
Changes in the \msigma\ relation at low mass translate directly into
uncertainties in the local BH mass function for \mbh$\lesssim
10^7$~\msun.  Furthermore, they imply ambiguity in the overall
scaling of BH masses in active galaxies based on reverberation
mapping.
\end{enumerate}

We now indulge in speculation about the physical ramifications of our
results.  Generic arguments about BH self-regulation, taken at face
value, seem to predict a simple power-law relationship between \mbh\
and \sigmastar\ \citep[e.g.,][]{silkrees1998,
  murrayetal2005,hopkinsetal2006}, although the shape and scatter of
the relation may depend on \mbh\
\citep[e.g.,][]{robertsonetal2006,crotonetal2006,debuhretal2010}.
This relation is established when the BHs grow large enough that their
accretion power can overcome the galaxy potential.  On the other hand,
the maser galaxies have systematically lower BH masses at a fixed
\sigmastar.  Perhaps because of the quiescent history of these spiral
galaxies, there has never been a major feeding episode in the past
history of their BHs.  If low-mass galaxies do not feed their BHs very
effectively, then the BH never approaches the ``limiting'' value set
by the galaxy potential \citep{hu2008,greeneho2008,youngeretal2008}.
Presumably the same processes that make a galaxy into an elliptical or
spiral (merging history) also determine the BH feeding efficiency.  In
this scenario, we expect a measurable difference in scatter between
elliptical and spiral galaxies with \mbh$\approx 10^7$~\msun.  We note
that ongoing adaptive-optics--assisted integral-field unit
observations are exploring the low-mass regime for elliptical galaxies
with improved statistics \citep{krajnovicetal2009}, which will allow
us to perform this test.

Alternatively, \citet{peng2007} proposes that BH-bulge scaling
relations are tightened at high mass due to multiple minor mergers,
which, by the central limit theorem, yield a tight \msigma\ relation
at the present day \citep[see
also][]{hirschmannetal2010,jahnkemaccio2010}.  In this picture the
only important parameter is number of mergers, rather than detailed
galaxy morphology.  Here we might expect to find increased scatter in
both the elliptical and spiral galaxies at \mbh$\lesssim 10^7$~\msun.
It is interesting to note, in passing, that one might also expect
globular cluster frequency to correlate more tightly with \mbh\ than
bulge luminosity \citep{burkerttremaine2010}.  Furthermore, we might
also expect this process to lower \mbh/$M_{\rm bulge}$ over cosmic time
\citep[e.g.,][]{greeneetal2010}.

This paper represents an initial exploration of scaling relations for
the maser galaxies, but there are many experiments we would like to
perform.  As noted above, further observations are required to
determine whether or not there are small classical bulges embedded in
all of these galaxies, whose \sigmastar\ and $M^*$ correlate tightly
with \mbh\ \citep{nowaketal2010}.  Integral-field spectroscopy would
reveal the \sigmastar\ profile and allow us to determine whether there
is a transition scale at which we recover a tight correlation between \mbh\
and \sigmastar.  Multi-color photometry, ideally with \hst\ imaging of
the nuclei, will allow us to decompose the light profile into its
constituent parts, measure stellar masses for each, and then determine
whether the BH mass correlates best with any particular combination of
components.

There are other potentially interesting BH-bulge scaling relation
projections to investigate as well.  For instance, it may be that
\mbh\ correlates more tightly with the dynamical mass than \sigmastar\
or stellar mass \citep[e.g.,][]{haeringrix2004,hopkinsetal2007}.
There is also a suggestion in the literature that \mbh\ actually
correlates with the dark matter halo mass
\citep[e.g.,][]{ferrarese2002}, although there are clearly
complications in spiral galaxies \citep[e.g.,][]{ho2007}.  At present
only a couple of the galaxies have resolved rotation curves, but we
will obtain \ion{H}{1} observations of the entire sample with the EVLA
in order to investigate both the gas content and the \mbh-$v_c$
correlation in these galaxies.

Beyond the present work there are now many suggestions that the
\msigma\ relation may not be the unbroken, low-scatter power-law
originally proposed
\citep[e.g.,][]{gebhardtetal2000a,ferraresemerritt2000}.  With the
larger samples available to date, the scatter has increased by $\sim
0.1$ dex \citep{tremaineetal2002,gultekinetal2009}.  At the same time,
new analysis incorporating the impact of dark matter at large radius
\citep{gebhardtthomas2009} has shown that the BH masses in the most
massive ellipticals may be systematically underestimated by a factor
of $\sim 2$.  Finally, new dynamical models incorporating triaxiality
may well increase the scatter in BH mass at a fixed bulge mass as well
\citep[e.g.,][]{vandenboschdezeeuw2010}.  Given that there is no
universal \msigma\ relation, it may be time to revist the role the BHs
play in galaxy evolution.

\acknowledgements{Many people have contributed substantively to our 
thinking in the writing of this paper. We first thank the referee for 
a very prompt and thorough report. We gratefully acknowledge useful
conversations with A.~Barth, L.~Ho, J. Kormendy, G. van de Venn, and
N. Drory.  P. Erwin and J. Mulchaey both provided optical images of
various galaxies presented here and K.  G{\"u}ltekin assisted in our
implementation of his fitting methodology.  

This research has made use of the NASA/IPAC Extragalactic Database
(NED) which is operated by the Jet Propulsion Laboratory, California
Institute of Technology, under contract with the National Aeronautics
and Space Administration. We further acknowledge the usage of the 
HyperLeda database (http://leda.univ-lyon1.fr).

Funding for the SDSS and SDSS-II has been provided by the Alfred
P. Sloan Foundation, the Participating Institutions, the National
Science Foundation, the U.S. Department of Energy, the National
Aeronautics and Space Administration, the Japanese Monbukagakusho, the
Max Planck Society, and the Higher Education Funding Council for
England. The SDSS Web Site is http://www.sdss.org/.

The SDSS is managed by the Astrophysical Research Consortium for the
Participating Institutions. The Participating Institutions are the
American Museum of Natural History, Astrophysical Institute Potsdam,
University of Basel, University of Cambridge, Case Western Reserve
University, University of Chicago, Drexel University, Fermilab, the
Institute for Advanced Study, the Japan Participation Group, Johns
Hopkins University, the Joint Institute for Nuclear Astrophysics, the
Kavli Institute for Particle Astrophysics and Cosmology, the Korean
Scientist Group, the Chinese Academy of Sciences (LAMOST), Los Alamos
National Laboratory, the Max-Planck-Institute for Astronomy (MPIA),
the Max-Planck-Institute for Astrophysics (MPA), New Mexico State
University, Ohio State University, University of Pittsburgh,
University of Portsmouth, Princeton University, the United States
Naval Observatory, and the University of Washington.
}

\appendix

\vskip +0.5truein
\hskip 1.8in
\psfig{file=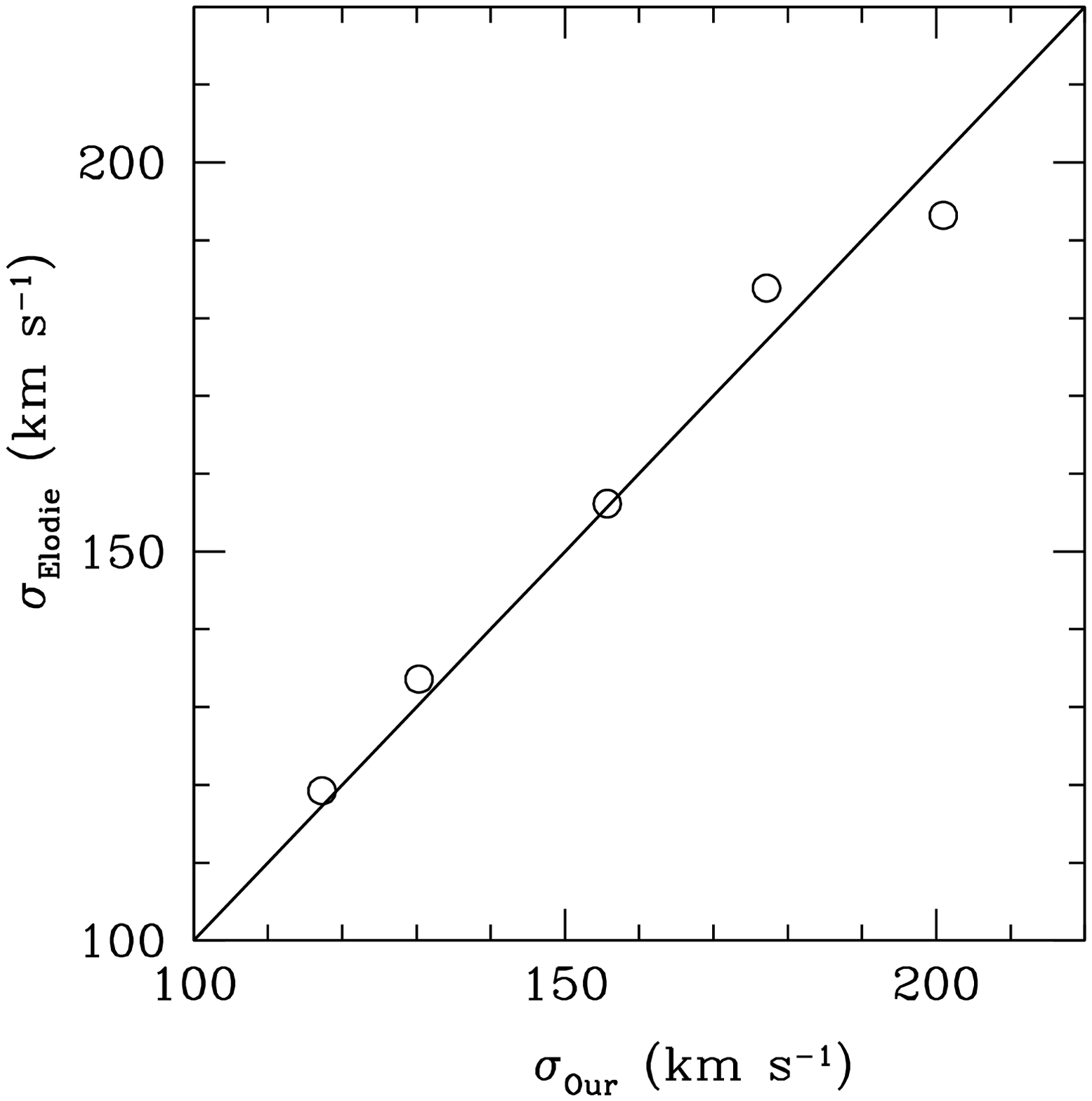,width=0.44\textwidth,keepaspectratio=true,angle=0}
\vskip -0mm
\figcaption[]{
Comparison of stellar velocity dispersion measurements using our
internal velocity template stars ($\sigma_{\rm Our}$) and the Elodie
template stars ($\sigma_{\rm Elodie}$) for the IMACS ({\it circles})
data.  The template stars, while not identical, are matched
approximately in spectral type to remove ambiguity caused by
template mismatch.  We measure the differential resolution by
fitting the internal template stars with the Elodie templates
\citep{moultakaetal2004} and use these values to correct the Elodie
fits for differing instrumental resolution.  For demonstration, two additional
galaxies are plotted here that are not in the current sample.
\label{cfsigma}}
\vskip 5mm

In addition to the B\&C data discussed in the primary text, we
obtained auxiliary spectra for some of the Southern targets in one
Magellan/Baade night.  We used the Inamori-Magellan Areal Camera \&
Spectrograph \citep[IMACS;][]{dressleretal2006} with the 1200
l~mm$^{-1}$ grating.  The primary setting had a tilt of 17.15 with a
resulting central wavelength of 4540~\AA, a spectral range of
3800-5300 \AA, and a dispersion of $\sigma_{\rm instr} \approx
15$~\kms.  We also obtained \cat\ observations for two targets, with a
tilt of 32.8, a resulting central wavelength of 8340~\AA, and a
spectral range of 7500-9000 \AA.  The observing conditions were not
optimal; it was only four days from full moon, and the typical seeing
ranged from 0\farcs7 to 1\farcs4 over the course of the night.

IMACS consists of an array of four by two CCDs with considerable tilt
in the spatial direction.  Therefore we used the COSMOS code developed
by G. Oemler, D. Kelson, and G. Walth to reduce long-slit IMACS data.
The COSMOS code has a variety of benefits over standard long-slit
reductions of the IMACS data.  For one thing, because the size of the
chip gaps is built into the code, we are able to solve for a global
(rather than a chip-by-chip) wavelength solution in a straightforward
manner.  Furthermore, the code contains an optical model for the
instrument and thus does a better job of modeling curvature in the
spatial direction.  We perform bias-subtraction, flat-fielding,
dispersion-correction, and rectification within COSMOS and then use
standard {\tt iraf} routines for tracing and extracting one-dimensional
spectra, including sky subtraction.  Finally, we use IDL routines as
described by \citet{mathesonetal2008} to perform flux calibration and
telluric absorption corrections.  

In addition to the IMACS spectra, three targets (NGC 1194, NGC 4388,
and NGC 6264) have spectra in the SDSS database.  These have a
spectral resolution of $\sim 70$~\kms\
\citep[e.g.,][]{greeneho2006sig} and completely cover the G-band
region.  We use Valdes template stars for the SDSS spectra as well as
the DIS spectra.

The velocity dispersion measurements proceeded as above, except that
our library of template stars taken with IMACS is limited to G and K
giants.  Thus we use a large library of template stars provided by the
Elodie catalog \citep{moultakaetal2004}.  These stars were observed
with $R \sim 42000$ or $\sim 3$~\kms\ resolution.  In all cases we
adopt the following stars as our templates: HD025457 (F5~V), HD148856
(G8~III), HD185351 (K0~III), and HD124897 (K1.5~III).

We fit our internal velocity dispersion templates with the Elodie
stars in order to determine the relative broadening between the two
systems.  In the case of IMACS the spectral resolution varies
significantly as a function of wavelength, but in the range $4100 <
\lambda < 4500$~\AA, fits to the IMACS stars with the Elodie templates
yield a relative dispersion of $12 \pm 0.5$~\kms.  As an additional
consistency check, we take our measured dispersions from the internal
velocity stars and compare them to those measured with the Elodie
stars in the identical wavelength range, where the latter are
corrected for the resolution difference in quadrature.  The result of
that test is shown in Figure 14.  The agreement is completely
satisfactory, with a fractional difference of $(\sigma_{\rm Our} -
\sigma_{\rm Elodie}) / \sigma_{\rm Our} = -0.02 \pm 0.03$.  We are
reassured both that there is no significant systematic offset between
the two sets of measurements and that the scatter is considerably
smaller than any other source of uncertainty.  

Overall, we find satisfactory agreement between the IMACS and B\&C
measurements (Table A5).  The one exception is the galaxy NGC 3393.
In this case, our IMACS data were not taken along the major axis, but
rather at PA=12$\degr$.  Since the IMACS measurement is nominally
consistent with our minor axis measurement, we suspect this PA
difference may at least partially explain the discrepancy.
Furthermore, we are able to test our flux calibration by comparing the
fluxes in \oiii\ between the two sets of observations.  Since we can
match approximately the slit size and position angle, any
discrepancies should predominantly reflect differences in flux
calibration.  It is reassuring that our measurements agree within
$\lesssim 30\%$ in all cases but NGC 2960, where the \oiii\ line falls
on a noisy part of the IMACS chip.

\vbox{ 
\vskip 0.2in
\hskip 2.8in
\psfig{file=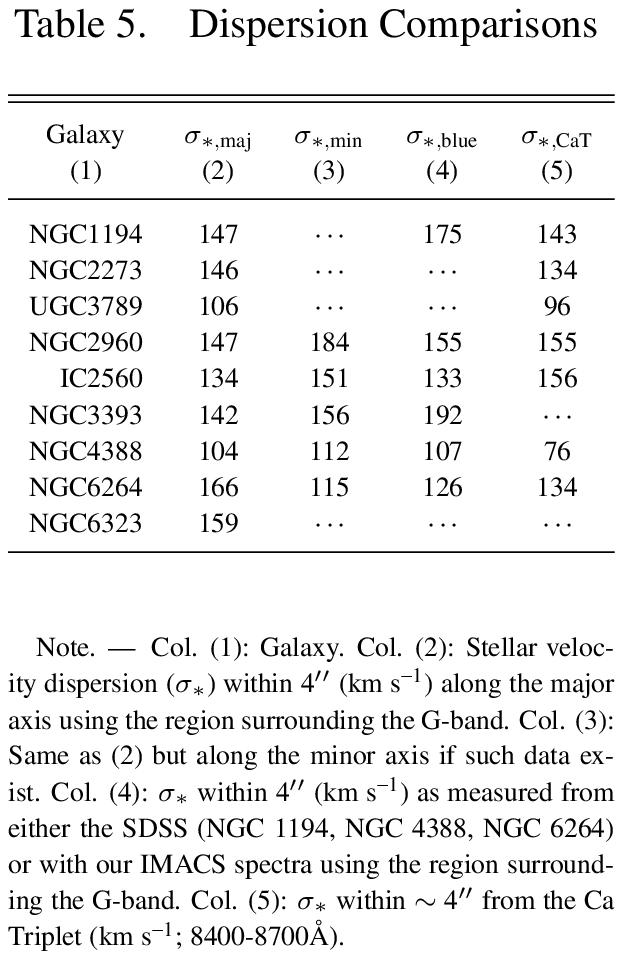,width=0.3\textwidth,keepaspectratio=true,angle=0}
}
\vskip 4mm

\end{document}